\documentclass[letterpaper]{article}%single
\usepackage[T1]{fontenc}

\usepackage{pdfpages}
\usepackage{geometry}
\usepackage{setspace}

\usepackage{achemso}
\usepackage{subfiles} %AZ for SI
\usepackage{graphicx}
\usepackage{float}
\newfloat{scheme}{htbp}{los}
\floatname{scheme}{Scheme}
\floatname{chart}{Chart}
\newfloat{graph}{htbp}{loh}

\usepackage{chemformula} % Formulas using \ch{}
\usepackage[version = 4]{mhchem} % Formulas using \ce{}
\usepackage{braket}%AZ
\usepackage{nicefrac}%AZ
\usepackage[dvipsnames]{xcolor} %AZ
\usepackage{braket} %AZ
\usepackage{amsmath} %AZ
\usepackage{booktabs} %AZ
\usepackage{float}
\usepackage{xfp}
\usepackage{threeparttable}
\usepackage{graphicx}
\usepackage{caption}
\usepackage{subcaption}
\usepackage{authblk}
\author[1]{Abdulrahman Y. Zamani*}
\author[1]{Kevin Carter-Fenk*}
\affil[1]{Department of Chemistry, University of Pittsburgh, Pittsburgh, Pennsylvania 15260, USA}

\title{Information-Theoretic Appraisal of Electron Densities}
\date{*Email: abdul.zamani@ronininstitute.org and kay.carter-fenk@pitt.edu}
\begin{document}

\maketitle

\begin{abstract}
We present an information-theoretic assessment of atomic and molecular densities in the ground state and under a range of physical scenarios---excitation, confinement, and ensemblization. Comparisons across densities obtained from single-reference methods are facilitated through information entropy measures evaluated in position space. We demonstrate that the J-divergence serves as a key metric for benchmarking electron densities against coupled cluster and configuration interaction references. Mean-field orbital information is further compared with that of Brueckner and Dyson orbitals, and informational changes in multiple self-consistent-field solutions are examined under various symmetry-breaking conditions. We also explore the relationship between entropic measures of electron delocalization and the accuracy of the CO dipole moment computed with different methods. Our work offers insights into the selection of optimal reference determinants for a given chemical application and highlights potential benefits of incorporating information-entropy concepts in the development of new density functionals.
\end{abstract}

\section{Introduction}

The electron density is the principal source and carrier of information\cite{Nalewajski2011,Geerlings2011} from which chemical structure, properties, and reactivity may be ultimately determined.\cite{Ruedenberg1962,Bader1981,Cremer1984,Bader1985a,Bader1985b,Bader1991,Matta2002,Bader2005,Proft2014,Popelier2014,Koch2024} The density and its deformation by external perturbation correspond to physical observables, which are accessible through X-ray diffraction\cite{Compton1915,Gavin1966,Coppens1977,Coppens1992,Tsirelson1996,Koritsanszky2001,Miao2002,Chapman2011,Kasai2018,Gao2019,Gruene2021}, Compton 
scattering\cite{Cooper1985,Tanaka2001,Achmad2004,Yu2016,Hiraoka2017,Feng2019,Talmantaite2020,Qiao2021}, and positron annihilation\cite{Stewart1957,Reed1976,Alatalo1996,Brusa2002,Wagner2018,Arias2024} experiments. 

Since the exact density encodes the correct physics to describe the many-electron interactions of a system, there is a strong case for its use in extracting information beyond the calculation of total energy differences. There has been a gradual resurgence of information theory\cite{Shannon1948,Jaynes1957} for quantum chemistry\cite{Gadre2002,Nalewajski2006,Rong2020,Zhao2025} with applications to  spectroscopy\cite{gadre1979,Sears1981,Gadre1985,Parr1986,Ho1998a,Restrepo2015,Schattschneider2020,Lee2020,Aguiar2022},  bonding\cite{Nalewajski2000b,Nalewajski2003,Nalewajski2004a,Nalewajski2004b,Nalewajski2012,Wu2019,Pal2022},  reactivity\cite{Nalewajski2002a,Nalewajski2002b,Esquivel2009,Esquivel2010a,Welearegay2014,Liu2016,FloresGallegos2022,FloresGallegos2023a}, molecular similarity\cite{Lin1996,Ho1998b,Nalewajski2000a,Godden2000,Nalewajski2001,FloresGallegos2021,FloresGomez2024}, and the more general electron correlation problem.\cite{Lowdin1958,Wigner1934,Collins1993,HoSagar1994,Lowdin1995,Ziesche1995,Esquivel1996,Ramirez1997,GoriGiorgi2002,Gottlieb2005,Gottlieb2007,Mahajeri2009,Mohajeri2009,Chai2012,FloresGallegos2016,Zhou2016,Lin2017,Alipour2018,FloresGallegos2023b,Gibney2022,Chen2022,Yeh2022,Gibney2023,Gibney2024,Zamani2025} A formal description of electronic structure in terms of the electron distribution is admitted by density functional theory\cite{Hohenberg1964,Levy1979,Parr1982,Parr1995,Kohn1996,Burke2012} (DFT) and associated response theories,\cite{Runge1984,Casida1995,Petersilka1996} upon which many quantum chemistry methods have been developed. Within the Kohn-Sham approach,\cite{Kohn1965}  the relative performance of density functional approximations (DFAs) depends on specific design principles, such as those motivated by empirical parameterization\cite{Cohen2008,Cohen2012,Jones2015,Mardirossian2017,Martin2020} and exact constraints\cite{Burke1995,Perdew1996,Burke1997,Perdew1998,Perdew2005,Perdew2014,Sun2015,Perdew2016,Kaplan2023}.

The quality of electron densities input into wave function approaches, such as coupled-cluster theory
and third-order M{\o}ller-Plesset perturbation theory, can have dramatic impacts on calculated properties
from vibrational frequencies to potential energy surfaces.\cite{Rettig2020,Bertels2021,Zamani2026} % need to cite our work and MP3 and CC papers
Along the vein of DFA design, alternative densities may be implanted into DFA energy functionals as a diagnostic of
density-driven and/or functional-driven errors. Such approaches have recently inspired the use of so-called density-corrected DFT\cite{Becke1988,Verma2012dcdft,Kanungo2024} in condensed-phase simulation
and have been used to suppress delocalization errors that plague the many-body expansion.\cite{Dasgupta2021,Broderick2024,Zulueta2025} % cite DC-SCAN and John Herbert's SIE in the MBE paper
Clearly, the quality of density that is input into our approximate methods matters,
so it is only natural that we appraise densities derived from commonly-used DFAs,
especially since the quality of densities produced by modern DFAs has been called into question.\cite{Medvedev2017} % cite Science
In this work, we will examine the fidelity of many-electron densities obtained from a variety of DFAs with those calculated from correlated wavefunction and many-body methods.  

\section{Information Theory}

The electron density function $\rho(\textbf{r})$ defines the probability of finding any one of a system's $N$ electrons at a point $\textbf{r}$. The Shannon entropy $S_\rho$ in position-space is defined as 

\begin{align}
    S_\rho = -\int \rho(\textbf{r}) \ln \rho(\textbf{r}) d\textbf{r}
\end{align}
and is interpretable as measure of information when $\rho(\textbf{r})$ is normalized such that, 

\begin{align}
    N= \int \rho(\textbf{r}) d\textbf{r}. 
\end{align}

The object $\rho(\textbf{r})~N^{-1}$ is known as the shape function\cite{Parr1983,Ayers2006} and is sufficient to determine the value of any observable quantity.\cite{Ayers2000} The Shannon entropy is also  defined in momentum-space $\textbf{p}$

\begin{align}
    S_\gamma = -\int \gamma(\textbf{p}) \ln \gamma(\textbf{p}) d\textbf{p}.
\end{align}

The electron momentum density $\gamma(\textbf{p})$ is related to $ \rho(\textbf{r})$ by a Fourier transform of the wavefunction and exhibits a slower asymptotic decay\cite{Kimball1975,Kulkarni1994} ($\sim p^{-8}$) than its position-space compliment. The sum of $S_\rho$ and $S_\gamma$ yields the total Shannon entropy ($S_T$) and is bounded by the Bia{\l}ynicki-Birula-Mycielski entropic uncertainty relations\cite{BBM1975,Bialynicki1984} in three dimensions

\begin{align}
    S_\rho +  S_\gamma \geq 3(1+\ln\pi). 
\end{align}

This inequality is a stronger statement of uncertainty for the canonical conjugate spaces than the well-known Heisenberg principle. The following bound\cite{Bialynicki1984} depicts this more clearly

\begin{align}
    S_\rho +  S_\gamma \geq 1- \ln 2 - \ln\Bigg[\frac{\Delta 
    \mathrm{\textbf{x}} \ \Delta 
    \mathrm{\textbf{p}}}{h} \Bigg]. 
\end{align}

The Shannon entropy can be thought of as a global measure of uncertainty, information spread, or delocalization of the underlying probability distribution. Equipped with the entropic uncertainty relations, one can see that $S_\rho$ and $S_\gamma$ behave in a reciprocal manner. As $S_\rho$ increases, knowledge of $\rho(\textbf{r})$ is minimized, $\gamma(\textbf{r})$ ``sharpens'', and $S_\gamma$ decreases. Here, the response of $S_\gamma$ to $S_\rho$ is also reflected in the lowering of momentum or kinetic energy (a  so-called randomizing force \cite{Sears1980}) as electrons become more delocalized. This counterbalance effect embodies   the physical significance of information entropy in the quantum theory of many electrons. 

Futhermore, $S_T$ has been used to gauge wavefunction and basis set quality in atomic and molecular systems.\cite{Ho1994,HoSagar1994,Ho1998a2,Ho1998b,Ho1998c,Guevara2003,Lin2015} If the known constraints of the system are satisfied, increases in $S_T$ signify a manifestation of the maximum entropy principle\cite{Jaynes1957b,Jaynes1983} by which the best probability distribution, i.e.\ $\rho$, can be selected. This work will focus on information-theoretic quantities centered on $\rho(\textbf{r})$ as common molecular quantum chemistry programs primarily make use of the position-space basis set representation.\cite{note1}
 
Let us now turn to some entropic measures of similarity that can be used to compare electronic densities acquired from various approximate solutions to the many-electron Schr\"{o}dinger equation. Direct comparison of densities based on $S_\rho$ alone is limited since two different, but comparable, densities can have the similar values of $S_\rho$. The Kullback–Leibler divergence\cite{Kullback1951} (KLD) instead quantifies the distance, \textit{loosely speaking}, between statistical distributions. In terms of two one-electron probability densities $\rho_1(\textbf{r})$ and $\rho_2(\textbf{r})$ we have 

\begin{align}
\mathrm{KLD}(\rho_{\text{1}} \| \rho_{\text{2}}) &= \int \rho_{\text{1}}(\mathbf{r}) 
\ln \frac{\rho_{\text{1}}(\mathbf{r})}{\rho_{\text{2}}(\mathbf{r})} \, d\mathbf{r}. 
\end{align}
%KLD or cross-entropy is invariant to coordinate transformation, unlike S_rho
Also known as relative entropy or information gain, the KLD gives the average information to discriminate $\rho_1(\textbf{r})$ from $\rho_2(\textbf{r})$ in order to determine the extent of their (in)distinguishability\cite{Borgoo2011} and has been used to analyze atomic and molecular densities.\cite{Ho1998b,Sagar2002,Chatzisavvas2005,Sagar2008,Laguna2019,10.1063/1.5124244} Note that the logarithm factor, or relative surprisal, is weighted by $\rho_1(\textbf{r})$. This leads to an asymmetry in the KLD since its value will depend on which density is used as the reference distribution. This can be rectified through symmetrization: 

\begin{align}
J\text{-}\mathrm{D} = \mathrm{KLD}(\rho_{\text{1}} \| \rho_{\text{2}}) + \mathrm{KLD}(\rho_{\text{2}} \| \rho_{\text{1}}).  
\end{align} 
This is known as the Jeffreys divergence ($J\text{-}\mathrm{D}$).\cite{Jeffreys1946} While these quantities are not metrics in a strict geometric sense, as they do not satisfy the triangle inequality,\cite{Endres2003,Nielsen2020} we will show that the $J\text{-}\mathrm{D}$ is a viable utility for qualitative density appraisal.

We now bring attention to the gradient of the density $\nabla \rho(\textbf{r})$, which encodes how the density varies in space. It is central to generalized gradient approximations\cite{Perdew1996} (GGA) in DFT and an important component of the Fisher information\cite{Fisher1925}

\begin{align}
I_F = \int \frac{|\nabla \rho(\mathbf{r})|^2}{\rho(\mathbf{r})} \, d\mathbf{r}. 
\end{align}
Complementary to $S_\rho$, $I_F$ measures the narrowness or concentration of the electron distribution. $I_F$ is closely related to the kinetic energy and thus to development of kinetic energy functionals.\cite{Wang1998,Ayers2005,Ghiringhelli2010,Hamilton2010,Mi2023,Sears1980,Ludena1982,Nagy2013}  Consider the Weizs\"{a}cker kinetic energy,\cite{Weizsacker1935} $T_W$, which is constructed from the gradient of an $N$-normalized density $n(\textbf{r})$: 

\begin{align}
T_W  &= \frac{1}{8} \int \frac{|\nabla n(\mathbf{r})|^2}{n(\mathbf{r})} \, d\mathbf{r} 
\end{align}
It can be shown that  $I_F$ and $T_W$ are proportional

\begin{align}
T_W &= \frac{N}{8} I_F 
\end{align}
---highlighting yet another connection between information theory and quantum mechanics.

Finally, we will briefly survey the Fisher-Shannon complexity\cite{Romera2004,Sen2007,Angulo2008,lopezRosa2009,Esquivel2010b} (FSC) 
\begin{align}
   \mathrm{FSC} \equiv \frac{1}{3}I_F J_\rho
\end{align}
which involves the product of $I_F$ and the three-dimensional Shannon entropy power\cite{Dembo1991}
\begin{align}
    J_\rho = \frac{1}{2\pi e}e^{{2 S_\rho}/{3}}. 
\end{align}
$J_\rho$ is the variance of a Gaussian that shares the same entropy as $\rho(\textbf{r})$\cite{Stam1959,Verdu2006,Rioul2011} and serves as a geometric measure of information spread. FSC measures both local order (or sharpness) and spatial extent by incorporating the average structural features of the density. We will now apply the information-theoretic measures introduced here to an assortment of problems in chemical physics.

\section{Computational Details} 

Molecular structures are obtained from the NIST database\cite{Nist}, unless stated otherwise. Self-consistent-field (SCF) calculations, including restricted (R), unrestricted (U), real-generalized (G), and grand-canonical (gc) Hartree-Fock (HF), are performed with modified versions of Q-Chem 6.3\cite{qchem} and PySCF\cite{pyscf}.  Calculations with natural-orbital functional (NOF) methods are performed with PyNOF\cite{pynof} and DoNOF,\cite{Piris2021,Lew-yee2026} where PyNOF's backend uses computational kernels from Psi4.\cite{psi4,Psi4NumPy} Spin-projected unrestricted Hartree-Fock (SUHF) calculations are performed with the ExSCF program.\cite{ExSCF} Configuration interaction singles and doubles (CISD) and coupled cluster singles and doubles (CCSD) are performed with PySCF or Psi4. Selected configuration interaction (SCI), complete active space (CAS), Brueckner coupled cluster doubles (BCCD), and auxiliary second-order Green’s function (AGF2) calculations are also performed in PySCF. SCF and CCSD vibrational frequencies are obtained with Q-Chem. All electrons are correlated in post-SCF methods. Information-theoretic measures are computed with each respective package's default numerical integration settings. Densities are evaluated over Lebedev-Laikov (Q-Chem, PySCF) or Lebedev-Treutler (Psi4) grids. Q-Chem uses pruned standard quadrature grids, SG1 and SG2, for the DFAs employed here.\cite{Gill1993,Dasgupta2017} PySCF uses atomic grids set to include a range of radial (50-105) and angular (302-434) points. Psi4 uses a Lebedev-Treutler (75, 302) grid with Treutler atomic weight partitioning.\cite{Treutler1995}  Finally, the unrelaxed one-particle density matrix (ODM) is computed for each level of theory (relaxed ODMs are not used).

\section{Results and Discussion} 

In addition to surveying the performance of Hartree-Fock methods and common low-rung DFAs, we include the family of QTP functionals designed to improve the accuracy of spectroscopic properties derived from KS orbitals. The QTP functionals operate under the ionization potential (IP) theorem\cite{Bartlett2005} and related conditions\cite{Windom2022} for physically motivated parameterization of the exchange-correlation functional---leading to accurate results for charged and neutral excitations involving valence and inner-shell electrons.\cite{Verma2012,Verma2014,Jin2018,Ranasinghe2019} As an additional effect of correcting the KS orbital energies, self-interaction error\cite{PerdewSIE} is shown to be greatly diminished.\cite{Ranasinghe2017} Piris' global natural orbital functional\cite{Piris2021gnof} (GNOF) method is selected for its balanced description of dynamic and non-dynamic electronic correlation through accurate reconstruction the two-particle reduced density matrix. Charge delocalization error is also significantly reduced with GNOF.\cite{LewYee2022,LewYee2023} To simulate correlated or ensemble densities with a single-reference determinant, we employ a gcHF method akin to recently developed information-theoretic mean-field methods.\cite{Irimia2023} 

\subsection{Ground State Information}
\begin{figure}[H]
    \centering
    \includegraphics[width=0.95\linewidth]{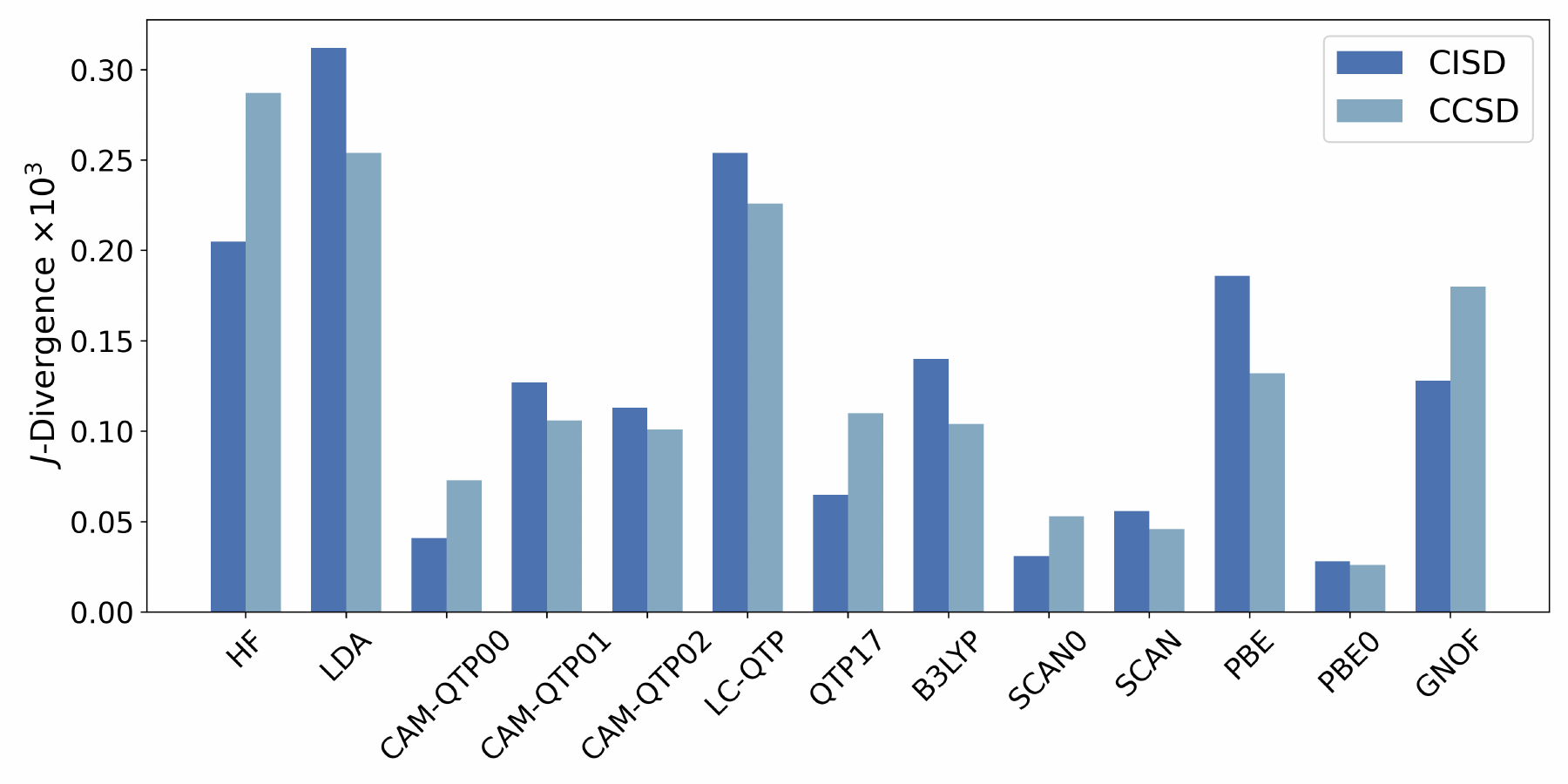}
    \caption{Jeffreys divergence with respect to CISD and CCSD densities for \ce{H_2O} at the NIST experimental geometry. Results are obtained with the cc-pVTZ basis set.}
    \label{fig:h2o}
\end{figure}

\begin{figure}[H]
    \centering
    \includegraphics[width=0.95\linewidth]{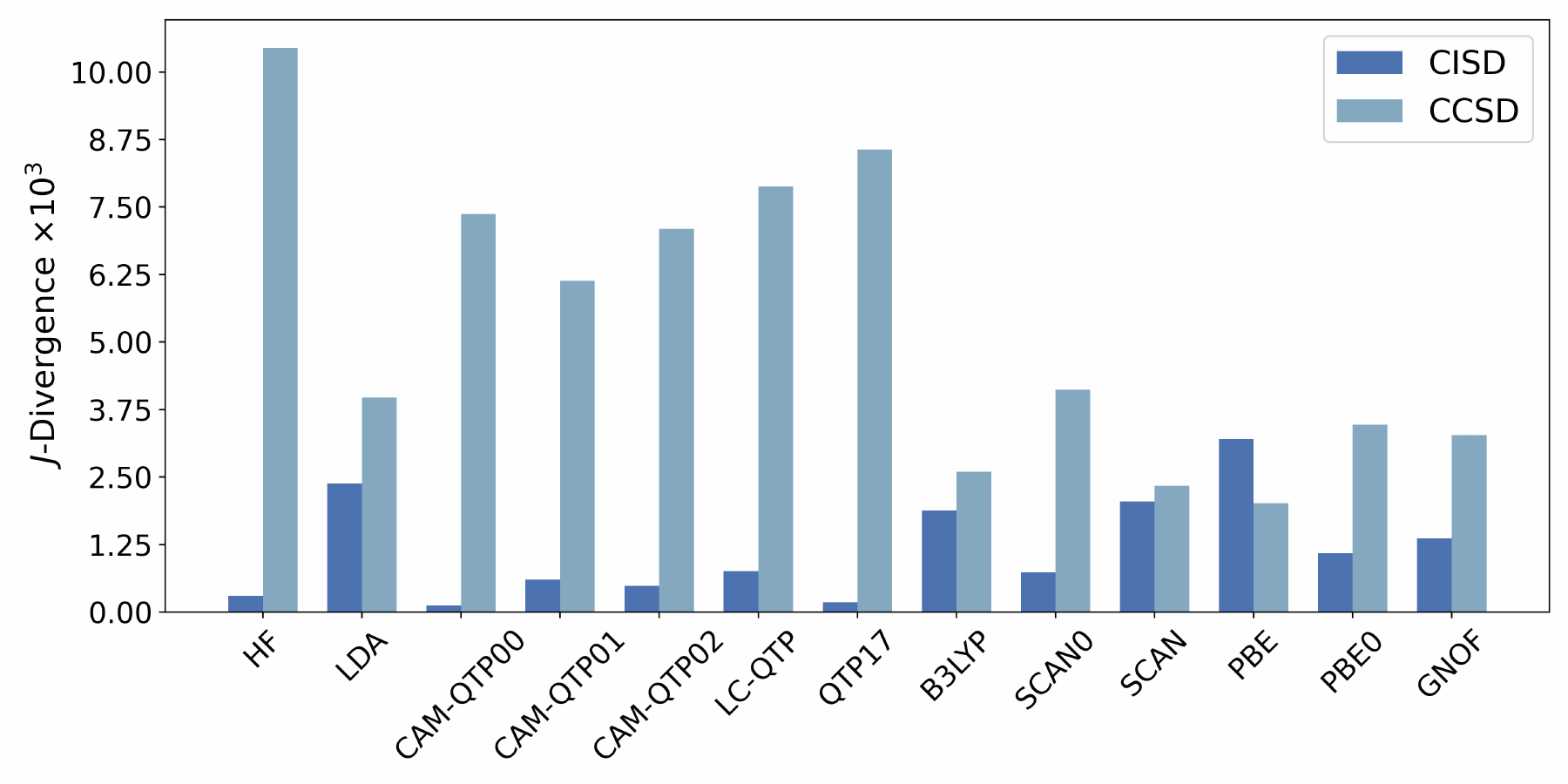}
    \caption{Jeffreys divergence with respect to CISD and CCSD densities for \ce{(H_2O)_4} at the geometry obtained from the WATER27\cite{Bryantsev2009} subset of GMTKN55.\cite{Goerigk2017} Results are obtained with the aug-cc-pVDZ basis set.}
    \label{fig:h2o-4}
\end{figure}

\begin{figure}[H]
    \centering
    \includegraphics[width=0.95\linewidth]{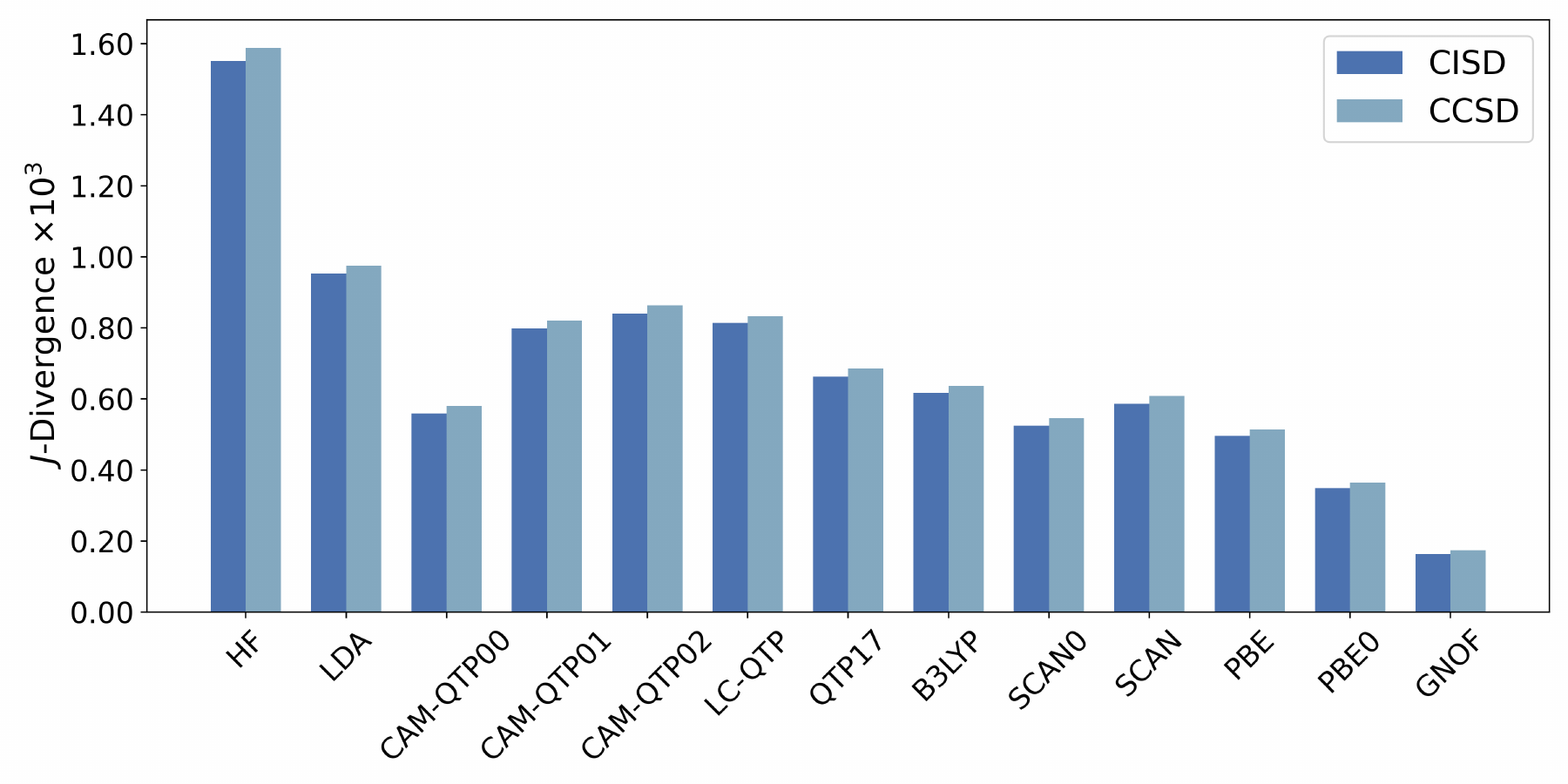}
    \caption{Jeffreys divergence with respect to CISD and CCSD densities for \ce{Be}. Results are obtained with the cc-pVTZ basis set.}
    \label{fig:be-atom}
\end{figure}

\begin{figure}[H]
    \centering
    \includegraphics[width=0.95\linewidth]{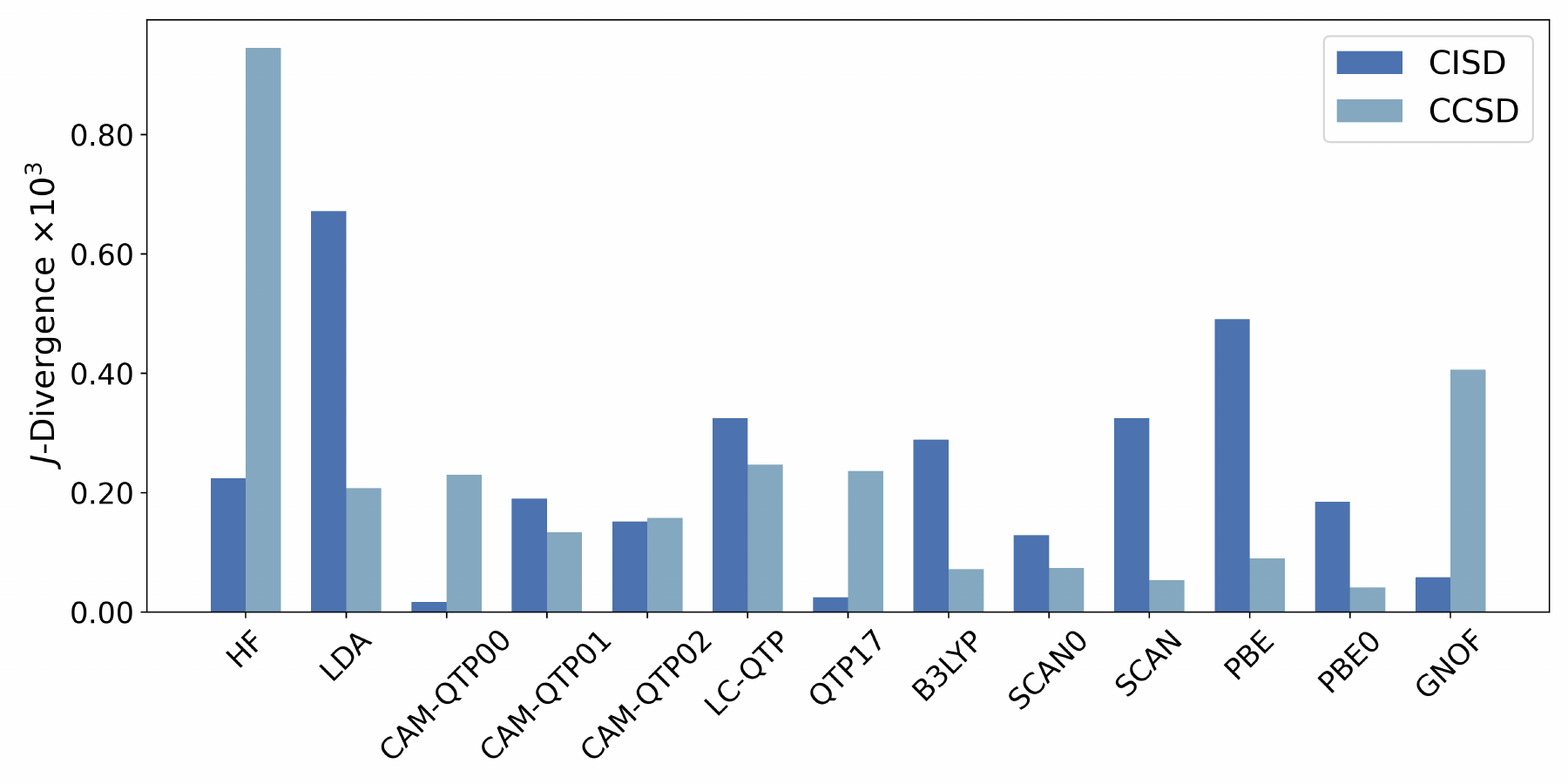}
    \caption{Jeffreys divergence with respect to CISD and CCSD densities for \ce{O_3} at the NIST experimental geometry. Results are obtained with the cc-pVTZ basis set.}
    \label{fig:o3}
\end{figure}

Differences between ground state CISD and CCSD densities and those calculated with HF, assorted DFAs, and GNOF are measured with the $J$-divergence. Results for \ce{H_2O}, \ce{(H_2O)_4}, beryllium atom, and \ce{O_3}, each with their own types of correlation effects, are displayed in Figures \ref{fig:h2o}-\ref{fig:o3}. Comparisons with SCI for \ce{Be} and \ce{H_2O} are displayed in Supporting Figures 1 and 2.

For single molecule \ce{H_2O} (Figure \ref{fig:h2o}), we see that the majority of DFAs and GNOF yield a $\rho(\textbf{r})$ more similar to CISD or CCSD than HF or the local density approximation (LDA). It is unsurprising that CAM-QTP00, a reparameterization of CAM-B3LYP to satisfy the IP theorem for the occupied
orbitals of \ce{H_2O}, provides a good-quality density. On the other hand, the LC-QTP\cite{Haiduke2018} density is appraised to bear little difference from HF or LDA densities. The other QTP methods, also tuned for improved orbital eigenvalues across a broad set of systems and properties (charge-transfer, electron attachment, etc.) yield smaller $J$-D values, but still offer improvements over a HF reference. Close in performance to CAM-QTP00 is the SCAN meta-GGA and its hybrid variant SCAN0, along with PBE0. Notably, SCAN and PBE are non-empirical functionals that are parameterized to satisfy exact constraints on the electron density, though HF exchange appears to be an important factor in the quality of density produced by the latter. The results with SCAN in particular appear to justify its previous applications to water simulation\cite{Dasgupta2022,Song2023} when supplemented by density corrections.

Results for the water tetramer \ce{(H_2O)_4} (Figure \ref{fig:h2o-4}), indicate that all methods generally produce densities much closer to CISD than CCSD. The $J$-divergence between CISD and CCSD is also exceptionally large for \ce{(H_2O)_4} compared to the other species examined in this section (see Supporting Table 1). This disparity on the side of CC underpins the need to include correlation effects to describe the non-covalent interactions in water clusters of increasing size,\cite{Howard2014} and might manifest due to size-extensivity errors in CISD. GNOF and the non-QTP DFAs seem to offer more balanced ground state densities for  \ce{(H_2O)_4}. Although the QTP functionals were developed with a focus on refining the KS potential over reconstruction of a correlated density, they again improve upon HF. %and offer additional utility in computational spectroscopy. 
We also note that the performance of SCAN and PBE in reproducing both CISD and CCSD correlated densities appears to be exceptional for \ce{(H_2O)_4}.

Beryllium is a prototypical atomic system used in benchmarking studies on highly accurate wavefunction methods.\cite{Esquivel1987, Finley1996, Bunge2009, Hornyak2019} In Figure \ref{fig:be-atom}, we see that non-HF $J$-divergences are $39-89\%$ smaller than the HF measure. The corresponding SCF densities are also are comparatively similar relative to CISD or CCSD. In particular, the hybrid flavors of SCAN and PBE, along with CAM-QTP00 and GNOF, show that much less information is lost when their SCF densities are used to estimate the correlated references. Out of these methods, the $J$-divergence suggests that GNOF offers the best approximation of the exact $\rho(\textbf{r})$.

Finally, in Figure \ref{fig:o3}, we appraise densities for singlet ozone, which is known to possess appreciable multi-reference character in the ground state. There is evidence to suggest that \ce{O_3} can be treated with single-reference DFAs, such as SCAN, specifically in the context of X-ray spectroscopy.\cite{Hait2024} Relative to HF, the densities are improved with most of the methods considered. It is noteworthy that the QTP functionals, despite not being parameterized for multi-reference systems, do sufficiently well for \ce{O_3}. The $J$-divergence comparing $\rho^\mathrm{CISD}$ and $\rho^\mathrm{CCSD}$ for \ce{O_3} is an order of magnitude less than for \ce{(H_2O)_4}, but it is not vanishing. This is likely reflected in the general trend of the densities being more similar to one over the other. Although obtaining accurate equilibrium properties of \ce{O_3} may present a challenge for some Piris functionals \cite{Mitxelena2016,Mitxelena2017}, GNOF has been shown to be capable of predicting the ionization potentials of extended \ce{O_3} chains.\cite{LewYee2023} As such, we find it reasonable to infer that close adherence to IP conditions and density matrix corrections can lead to some improvements in the quality of $\rho(\textbf{r})$, even in strongly correlated systems. 

For several methods in each example, the $J$-divergence results suggest greater similarity to one reference density over the other. Since CCSD captures higher-order correlation effects and is size-extensive, a density that more closely resembles a truncated CC reference is more reliable than one resembling a truncated CI reference at the same excitation order.
 
%AZ h2o-4 with hf is actually that good, doubled checked with diff basis and got same KLD

%cite piris o3 frequency paper, gnof was good

% CISD and CCSD are very different now that we have a water cluster

\subsection{Dissociation and Symmetry Breaking}

\begin{figure}[H]
    \centering
    \includegraphics[width=0.95\linewidth]{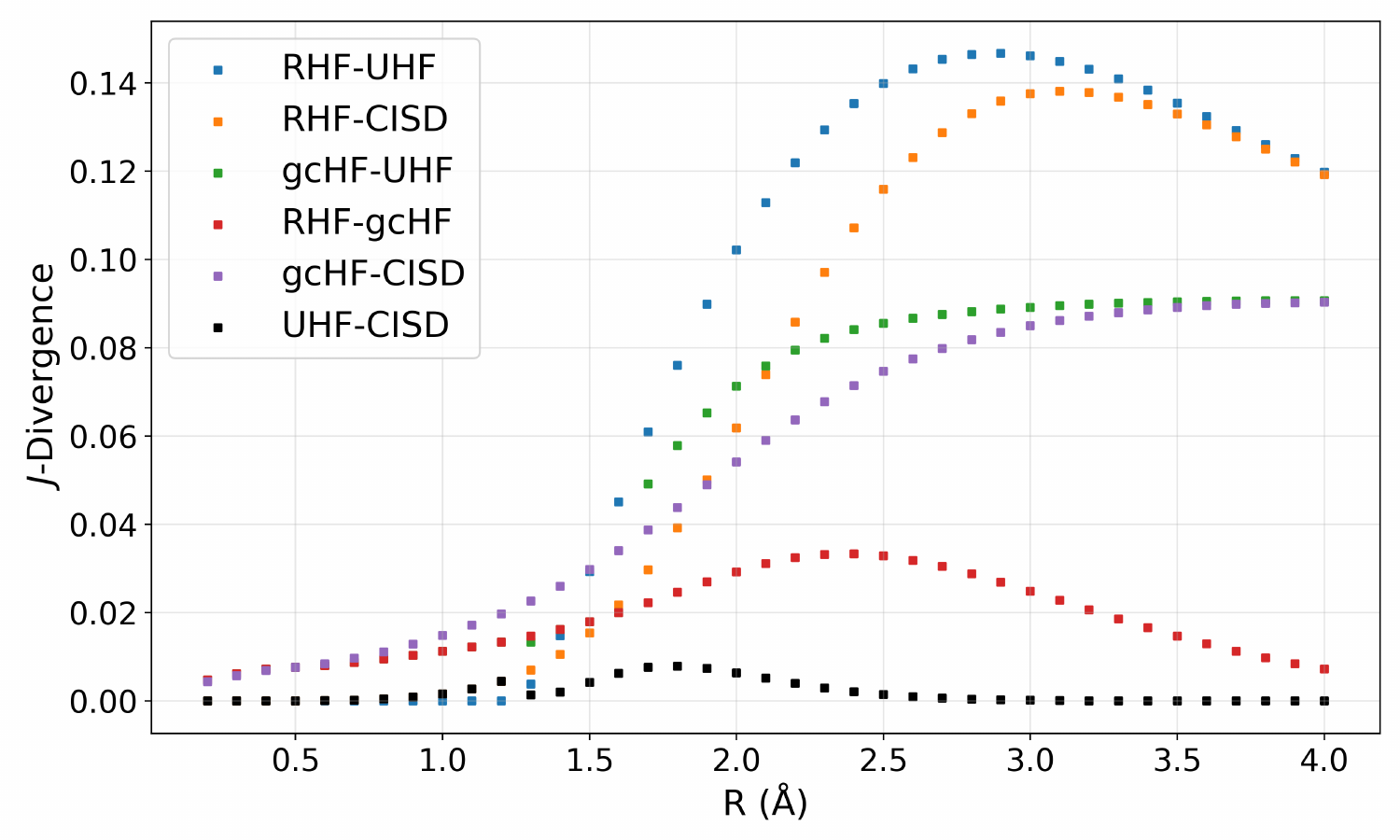}
    \caption{Jeffreys divergence relative to SCF and CISD densities for \ce{H_2} dissociation. Results are obtained with the cc-pVTZ basis set.}
    \label{fig:h2-stretch-JD}
\end{figure}

\begin{figure}[H]
    \centering
    \includegraphics[width=0.95\linewidth]{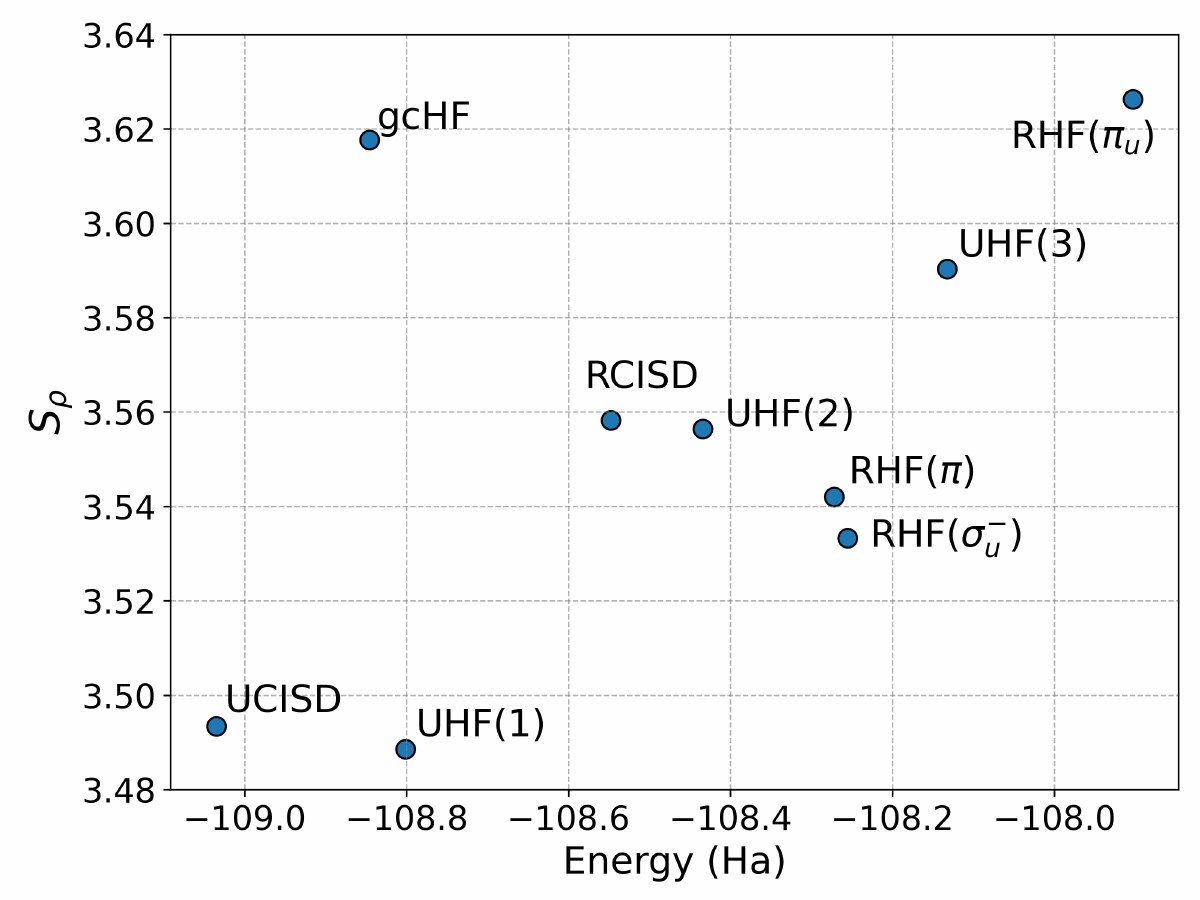}
    \caption{Shannon entropy for \ce{N_2} at 4\text{\AA} under various symmetry breaking conditions. Results are obtained with the cc-pVTZ basis set.}
    \label{fig:n2-symm}
\end{figure}

\begin{figure}[H]
    \centering
    \includegraphics[width=0.95\linewidth]{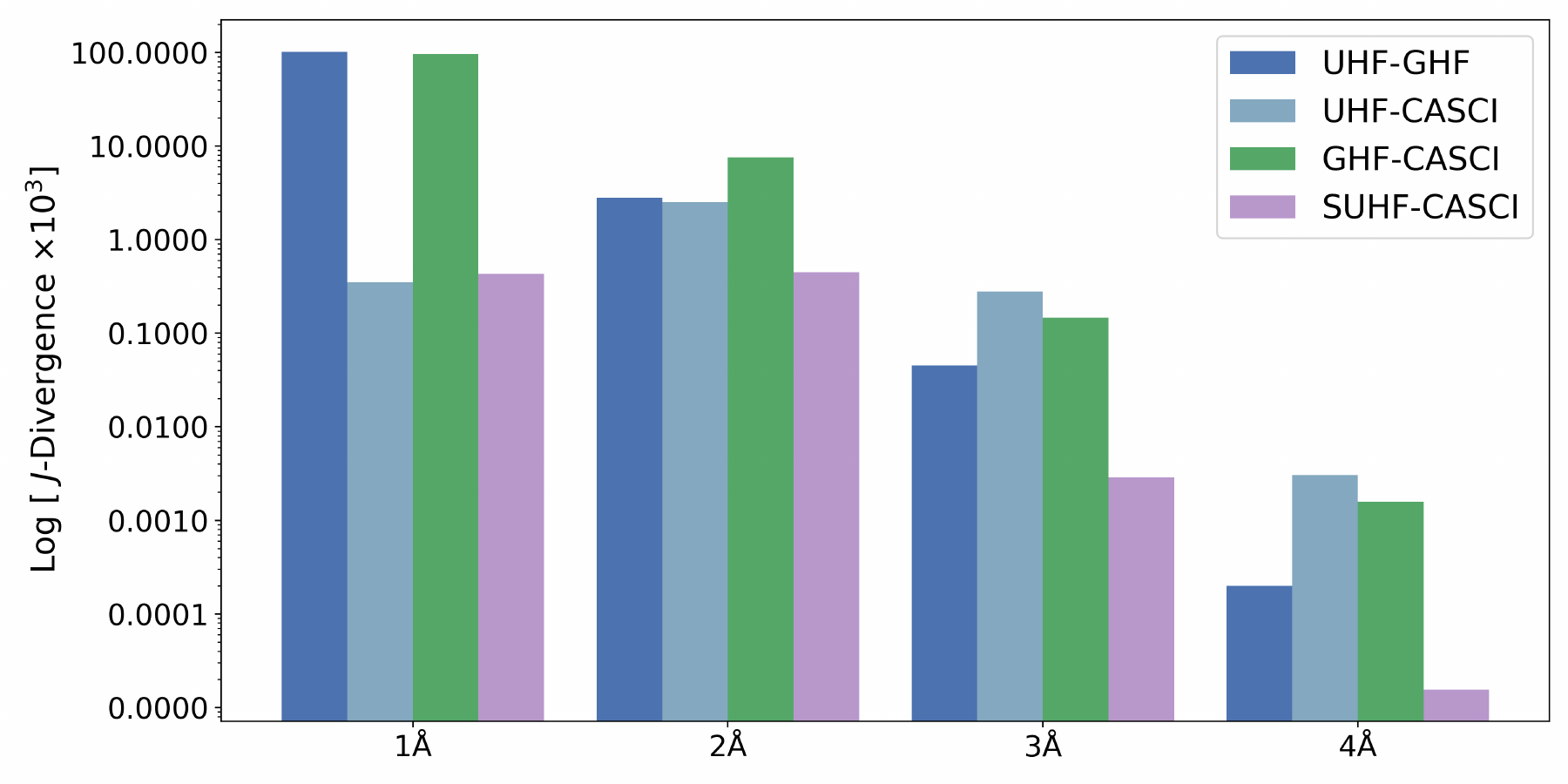}
    \caption{Logarithm of Jeffreys divergence for incrementally stretched \ce{H_3} under various symmetry breaking conditions. Results are obtained with the cc-pVTZ basis set.}
    \label{fig:h3-symm-JD}
\end{figure}

In this section we examine changes in relative information in response to symmetry breaking (Figures \ref{fig:h2-stretch-JD}-\ref{fig:h3-symm-JD}). Spin and spatial symmetry breaking can occur when the constraints on those degrees of freedom are lifted.\cite{Fukutome1981,Stuber2003} In the onset of bond dissociation, strong correlation effects arise, making it necessary to break the symmetry of the reference determinant to obtain qualitatively correct results. We compare HF methods against CISD in this scenario for \ce{H_2}. Note that despite size-extensivity errors in CISD, it is exact in two-electron systems and is therefore a suitable benchmark for \ce{H_2}.

Figure \ref{fig:h2-stretch-JD} displays $J$-divergences over stretched \ce{H_2} bond lengths and Supporting Figure 3 displays the corresponding potential energy curve. Here, the usual RHF-to-UHF triplet instability at the Coulson-Fischer point is indicative that a lower-energy, broken-symmetry solution exists. In Supporting Figure 4, see that $S_\rho$ for each method maximizes at dissociation. As expected, the RHF energy and density are quantitatively and qualitatively different than those belonging to CISD. The UHF-CISD $J$-divergence is minimal, since spin symmetry is sacrificed in return for a lower variational energy.

Additionally, one may lift the constraint of binary occupation numbers (ON) by stratifying electrons into virtual orbitals. This is achieved through an $N$-conserving grand-canonical or thermal HF approach, where the fractional ONs are obtained using the Fermi-Dirac distribution at a fixed chemical potential. As is typical with such approaches, electronic entropy or correlation corrections are modulated with a fictitious temperature parameter $\theta$ set to yield the correct mono-radical ONs when the frontier molecular orbitals (MO) of \ce{H_2} become degenerate. Thus, the approximate gcHF density matrix aims to capture the equi-ensemble picture of stretched \ce{H_2}. Interestingly, the gcHF-CISD and gcHF-UHF $J$-divergences lie nearly midway between the RHF-CISD and UHF-CISD measures. Additionally, the average charge distributions of gcHF and RHF are similar even though their underlying ODMs may be different. 

%kld is not a transitive measure of similarity, A=B B=C A!=C, no triangle inequality

Next, in Figure \ref{fig:n2-symm}, we measure the Shannon entropy of HF solutions under different symmetry constraints for the stretched nitrogen molecule. Note that multiple UHF and RHF solutions can exist for the same state.\cite{Pulay1990,Thom2008,Rishi2016,Pulay1990} We locate two spatial-symmetry-broken RHF solutions ($\pi$, $\sigma_u^{-}$) where the highest occupied orbitals begin to localize on the separated \ce{N} atoms. These solutions are energetically lower than the symmetry adapted ($\pi_u$) RHF solution where the orbitals delocalized over the atoms instead. Global uncertainty in the density, as indicated by $S_\rho$, is decreased in the RHF solutions as spatial symmetry is lowered. $S_\rho$ for gcHF is also reduced slightly with respect to RHF($\pi_u$). 

The multiple UHF solutions are distinguished by their total spin-squared expectation values $\braket{\mathcal{S}^2}$ and by visual inspection of the orbitals. The lowest energy solution UHF(1) for stretched \ce{N_2} has an $\braket{\mathcal{S}^2}$ of 3.0. The two sequentially higher solutions are UHF(2) and UHF(3) with $\braket{\mathcal{S}^2}$ values of 2.0 and 1.0, respectively. The entropic ordering of UHF solutions  behaves similarly to the RHF trend: $S_\rho$ decreases as we lift constraints on spin and spatial degrees of freedom. 
These observations echo the symmetry dilemma\cite{Lowdin1963} and indicate that variational alterations to the ODM structure due to changes in symmetry constraints can change the $\rho(\textbf{r})$ obtained from its diagonal.\cite{note3} From Figure \ref{fig:n2-symm}, we find that breaking symmetry in the single determinant guides us closer towards CI, both energetically and entropically. The position-space entropy is larger for symmetry adapted RHF than in CISD, but we can infer from numerical evidence\cite{Ho1998a2} that $S_T$ generally increases upon inclusion of correlation effects. 
%https://arxiv.org/pdf/2604.18081
We note that a recent study\cite{Diogo2026} found that, in a minimal basis, the symmetry-adapted RHF $S_\rho$ will eventually coalesce with the CI $S_\rho$ at infinite separation for \ce{H_2}. With larger basis sets, $S_\rho$ for \ce{H_2} and \ce{N_2} RHF solutions lie above CI results. 

%in nats they did: N Sσ −N log N
%STO6G:
% Shannon entropy (RHF) = 4.192627384360693
% Shannon entropy (RCISD) = 4.192624828544486
%cc-pvtz:
%Shannon entropy (RHF) = 5.33499537960087
%Shannon entropy (RCISD) = 4.836175963273261

Finally, in Figure \ref{fig:h3-symm-JD}, we examine  $J$-divergences of the eternal triangle\cite{Harris2008}, \ce{H_3}. Strong correlation that emerges from spin-frustration can be treated with GHF, where the constraint of spin collinearity is lifted. The log[$J$-D] measures, in reference to full-space CI densities, are calculated for symmetrically dissociating \ce{H_3}. Shannon entropy differences for \ce{H_3} are displayed in Supporting Figure 5. Around equilibrium ($\sim 1$~\AA), the HF methods fail to recover the bulk of the correlation energy\cite{JimenezHoyos2012}, with SUHF possessing the lowest $\Delta S_\rho$, followed by UHF, and GHF. At 2--4~\AA, the trend for UHF and GHF are reversed in $\Delta S_\rho$, gradually approaching the lower SUHF values. The log[$J$-D] results suggest that the  densities are still qualitatively different among the methods. At 4~\AA, log[$J$-D] values decrease significantly, and the SUHF density becomes the most similar to that of CI.  

%no need for gcHF-CISD, rhf gc are similar.

\subsection{Excitation}

In addition to ground state systems, we briefly probe the density information of excited states. In Table \ref{table:excite}, we compute information measures and transition energies of \ce{Li_2}. Excited state UHF solutions are obtained with the maximum overlap method\cite{Gilbert2008,Corzo2022} (MOM). One-shot and self-consistent spin projection are performed atop broken-symmetry UHF solutions. Spin projection improves upon UHF estimates of vertical transition energy; SUHF gives the best result relative to experiment. The $J$-divergence between SUHF(1) and CISD is smallest in the ground state, and similar to UHF in the excited state.

\begin{table}[htbp]
\renewcommand{\arraystretch}{1.75}

\centering
\begin{threeparttable}
 
\caption{Results for the $X^1\Sigma^+_g$ and $A^1\Sigma^+_u$ states of \ce{Li_2}.$^{a}$ Shannon entropies and Jeffreys divergences with respect to  CISD$^{b}$ are computed with UHF and SUHF. Vertical transition energies $T_e$ (eV) are also included.}
\label{table:excite}

\begin{tabular}{lccccc}
\hline
\ce{Li_2}& {$S_{\rho}^\mathrm{GS}$} & {$S_{\rho}^\mathrm{EX}$} & {$J$-D}$_\mathrm{GS}$$^c$  & {$J$-D}$_\mathrm{EX}$$^c$& $T_e$$^{d}$
 \\
\hline

{UHF} &  4.194 & 4.307  & 1.099 & 0.466 & 1.17 (-0.57)
       \\

{SUHF} & 4.188  &  4.319 & 2.119 & 3.050 & 1.64 (-0.11)
     \\

{SUHF(1)$^{e}$} & 4.179  & 4.307 & 0.244 & 0.466 & 1.57 (-0.17)
     \\
\hline
\end{tabular}
%Fock is build 1x in SUHF(1)

\begin{tablenotes}
\footnotesize
\item[$a$] NIST experimental geometry and the def2-TZVPD basis set are used.
\item[$b$] {$S_{\rho}^\mathrm{GS}$} and {$S_{\rho}^\mathrm{EX}$} for CISD are 4.189 and 4.306 respectively.
\item[$c$] $J\text{-}\mathrm{D}\times 10^3$
\item[$d$] Signed errors with respect to experiment\cite{Kaldor1991} in parenthesis.
\item[$e$] One-shot (1) spin-projection.
\end{tablenotes}

\end{threeparttable}
\end{table}
 
%AZ looks like full SCF delta SUHF rho is further from CISD
%AZ PUHF from PMP2 in GDV is similar to doing 1shot SUHF
%we also break GS symm with SUHF by using a open-shell fragment guess
%1shot suhf doubled checked, SUHF gives same Srho,J-D but different Energy 
 
\subsection{Core-orbitals}

\begin{figure}[H]
    \centering
    \includegraphics[width=0.475\linewidth]{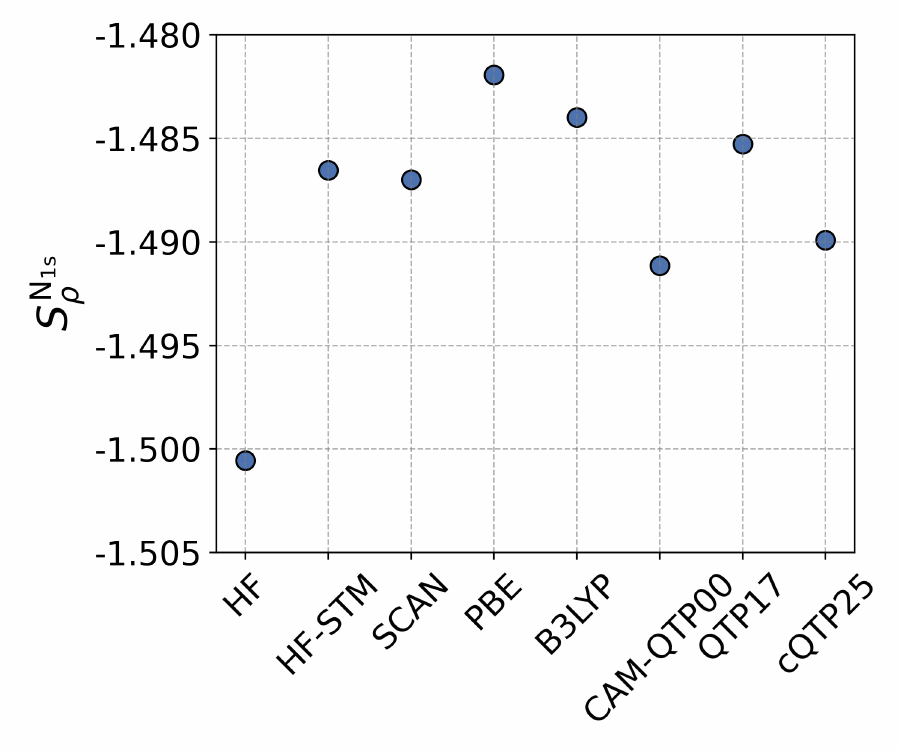}
    \includegraphics[width=0.475\linewidth]{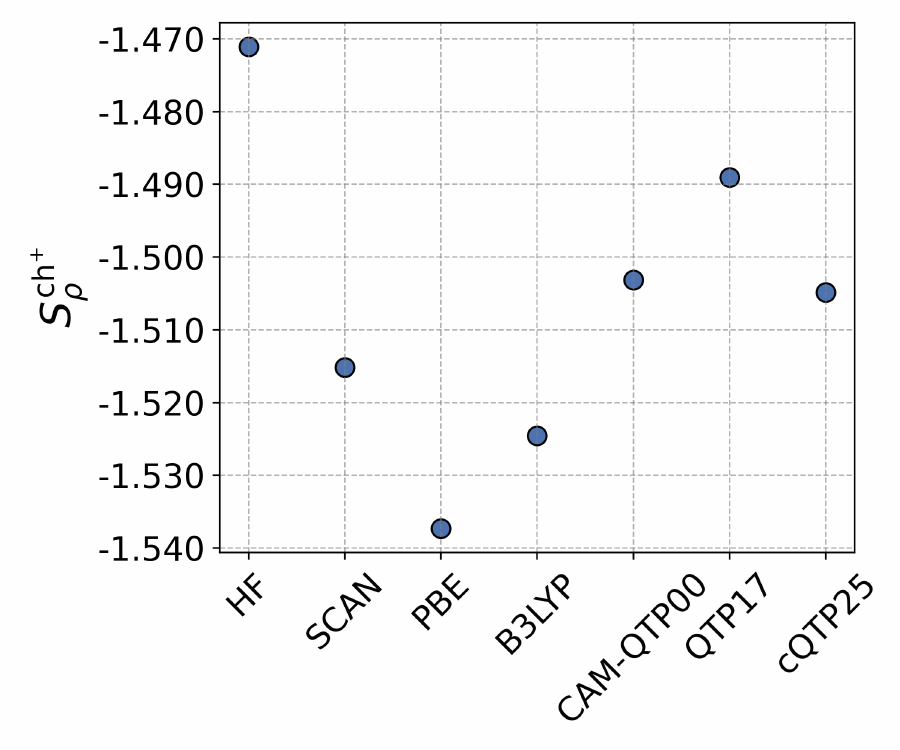}
    \caption{Shannon entropies for neutral core $S_{\rho}^\mathrm{N_{1s}}$ and core-hole $S_{\rho}^{\mathrm{ch^{+}}}$ orbitals of \ce{NH_3} obtained at the  calculated (CCSD(T)/cc-pVTZ) NIST geometry. The basis sets cc-pCVTZ and cc-pVTZ are assigned to \ce{N} and \ce{H} respectively.}
    \label{fig:nh3-1s}
\end{figure}

So far, we have surveyed information of the total density. The quality and shape of orbital densities have great importance, particularly in the one-electron picture of chemical bonding. We begin by inspecting the delocalization of core orbital densities. In Figure \ref{fig:nh3-1s}, $S_\rho$ is computed for the 1s orbital of \ce{NH_3} with different SCF methods. The   cQTP25\cite{Mendes2025} functional is included since it provides accurate core-electron ionization energies. The Slater transition method\cite{Slater1972} (STM) is used in tandem with a fractional occupation SCF solver to also optimize the core orbital. Orbital entropies are calculated from core-hole (ch$^+$) cation solutions obtained with MOM-SCF. The $S_\rho^\mathrm{N_{1s}}$ and  $S_\rho^\mathrm{ch^{+}}$ measures correspond to the occupied core 1s of the neutral parent molecule and the unoccupied core-hole, respectively. While we note that $S_\rho$ may be redefined to provide strictly positive values,\cite{FloresGallegos2019} we see that negative-valued $S_\rho$ are typical for tightly bound distributions (see Supporting Table 2). The $S_\rho^\mathrm{N_{1s}}$ value indicates that HF overly localizes the occupied 1s, whereas $S_\rho^\mathrm{ch^{+}}$ indicates that HF greatly delocalizes the core hole. In contrast, the tested DFAs act oppositely in each of the prior scenarios, delocalizing the occupied 1s orbital and overly-localizing the core hole. As Koopmans-like interpretations\cite{Koopmans1934} of the KS core-orbital energies supplied by QTP functionals are restored, the stark contrast between the $S_\rho^\mathrm{N_{1s}}$ values for the KS QTP orbitals and $S_\rho^\mathrm{N_{1s}}$ for the HF orbital hints that the density may be improved jointly with improvements to the ionization energy. This hypothesis will be corroborated in the section on Dyson orbitals.  

\subsection{Confinement}

\begin{figure}[H]
    \centering
    \includegraphics[width=0.475\linewidth]{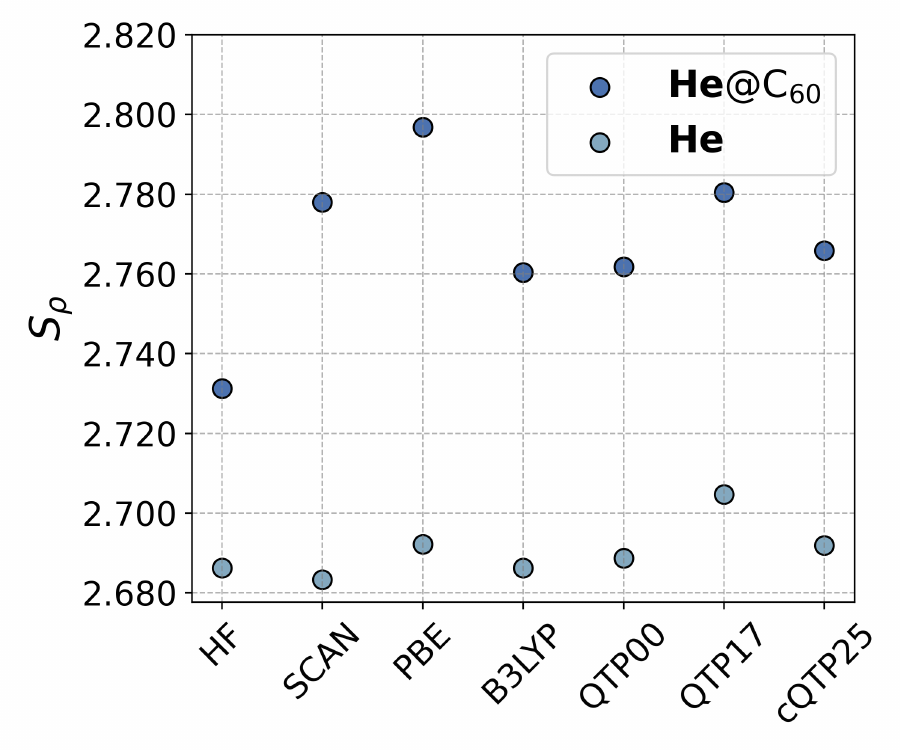}
    \includegraphics[width=0.475\linewidth]{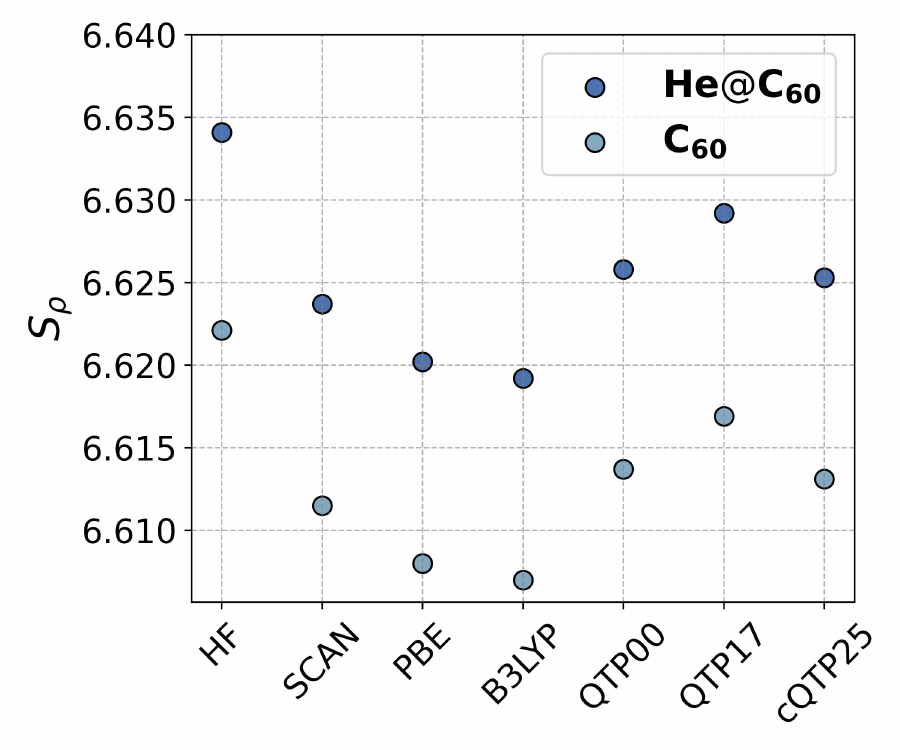}
    \caption{Shannon entropies for free \ce{He}, confined \ce{He}, free \ce{C_60}, and  \ce{He}@\ce{C_60}. Analyzed species are in boldface. The structure for \ce{C60} was optimized with PBEh-3c/def2-mSVP on a pruned (99, 590) grid. \cite{Dasgupta2017} Single-point calculations are performed with the cc-pVDZ basis set.}
    \label{fig:he-c60-confined}
\end{figure}

Molecules subject to spatial confinement can experience significant changes in bonding, electronic, and magnetic properties. For example, endohedral metallofullerenes\cite{Mansikkamaki2017,Popov2013} are highly attractive systems for quantum information processing,\cite{Atzori2019} as the fullerene cage acts as a confinement vessel. As an illustrative case, we analyze the endohedral confinement of helium atom through the Shannon entropy. Prior studies have modeled entropic changes of atomic and molecular systems in fullerene-like cavities and confining potentials.\cite{Mitnik2008,RodriguezBautista2018,GarciaMiranda2021,Mondal2023}
In Figure \ref{fig:he-c60-confined}, we compute the $S_\rho$ for free and confined \ce{He} in addition to the \ce{He}@\ce{C_60} complex and free \ce{C_60}. The increase in magnitude of $S_\rho$ for confined \ce{He} suggests that the 1s density becomes more delocalized relative to the free atom. This delocalization effect in confined \ce{He} is most prominent with SCAN, PBE, and QTP17. The relative trends of $S_\rho$ across methods also appear to be replicated for each species. Notions of entropy increases in ground state atoms upon confinement may not necessarily extend to excited states or different confinement radii where the delocalization behavior becomes more nuanced.\cite{Mondal2023,MartinezFlores2021}

%Srho goes up at 6bohr ~ c60 center to carbon radius, but can go down if confined further 
%
%https://doi.org/10.1002%2Fqua.25571

\subsection{Model Ensembleization}

We now consider entropic changes in systems with a degenerate manifold of states. Ensemble ODMs are obtained from CASSCF and gcHF calculations performed over windows containing degenerate orbitals. Information measures for p-block atoms are displayed in Table~\ref{table:smearing}. Configuration mixing, necessary for recovering non-dynamic correlation, is implicit in the single-particle weights---the ONs. This can be thought of as an enumeration of microstates. The resultant equi-ensembles where all ONs in the active space are equal, show an increase in information spread relative to the uncorrelated $S_\rho^\mathrm{UHF}$. We see that the symmetric divergences between gcHF and CASSCF densities are reduced significantly as more of the previously unoccupied single particle states become equally populated. This, in a way, mimics the delocalization of electrons at finite‑temperature and increases in uncertainty are reflected in the further maximization of $S_\rho$.

%fractionalization stratefied over fixed orbital windows
\begin{table}[H]%[htbp]
\renewcommand{\arraystretch}{1.75}
\centering
\begin{threeparttable}
\caption{Information measures for \ce{B}, \ce{C}, \ce{N},  and \ce{N^{+}} calculated using gcHF and state-averaged CASSCF within fixed orbital windows.$^a$ The cc-pVTZ basis set is used.}
\label{table:smearing}
\begin{tabular}{lcccc|cccccc}
\hline
& $(e,o)$$^b$ & {$S_{\rho}^\mathrm{gcHF}$} $^c$& {$S_{\rho}^\mathrm{CAS}$} $^d$&  {$J$-D}$^e$
& & $(e,o)$$^b$   & {$S_{\rho}^\mathrm{gcHF}$} $^c$& {$S_{\rho}^\mathrm{CAS}$} $^d$&  {$J$-D}$^e$ \\
\hline

{B} & (1,3) & 3.510 & 3.404 &    8.730 
    &  {B} & (1,6) & 3.620 & 3.610    & 0.077 \\

{C} & (2,3) & 3.163 & 3.103 &    2.125
    &  {C} & (2,6) & 3.447 & 3.436    & 0.086 \\

{\ce{N+}} & (2,3) & 2.404 & 2.360   & 1.167
    &  {N} & (3,6) & 3.231 & 3.221 & 0.056 \\
\hline
\end{tabular}

\begin{tablenotes}
\footnotesize
\item[$a$] $S_\rho^\mathrm{UHF}$ for \ce{B}, \ce{C}, \ce{N},  and \ce{N^{+}} are 3.373, 3.083, 2.796, and 2.343, respectively.
\item[$b$] (e)lectrons, (o)rbitals 
\item[$c$] Correlation parameter $\theta$ is tuned yield equal fractional occupations over the active space. 
\item[$d$]The maximum number of CASSCF roots is selected for state-averaging (equi-ensemble).
\item[$e$] $J\text{-}\mathrm{D}\times 10^3$
\end{tablenotes}
\end{threeparttable}
\end{table}

\subsection{Dyson Orbitals}

\begin{figure}[H]
    \centering
    \begin{subfigure}[b]{0.95\textwidth}
        \centering
        \includegraphics[width=\textwidth]{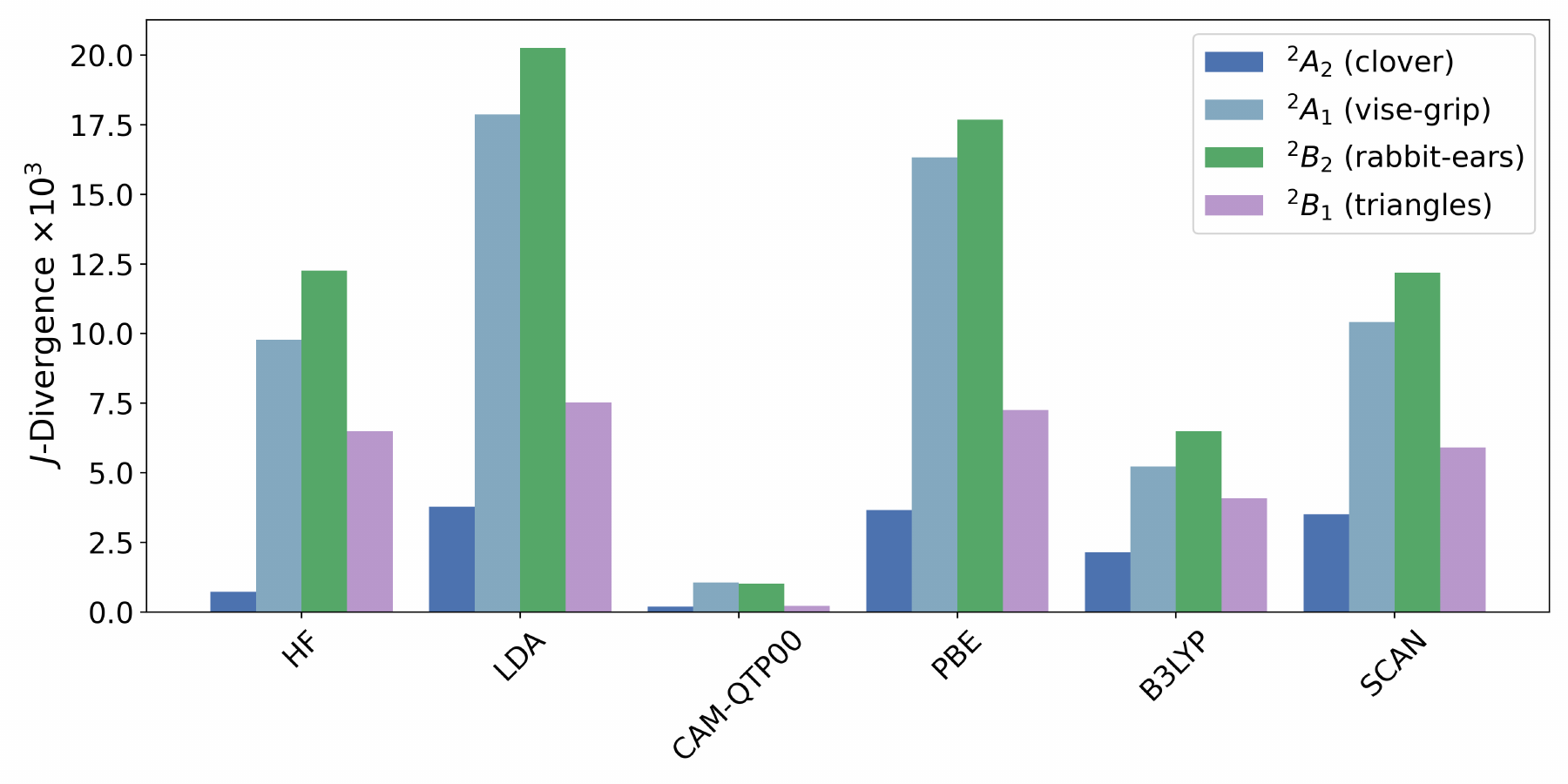}
        \caption{Jeffreys divergence with respect to AGF2 Dyson orbitals for \ce{O_3} at the NIST experimental geometry. Results are obtained with the cc-pVTZ basis set.}
        \label{fig:o3-dyson-JD}
    \end{subfigure}
    \vspace{0.5cm}
    \begin{subfigure}[b]{0.95\textwidth}
        \centering
        \includegraphics[width=\textwidth]{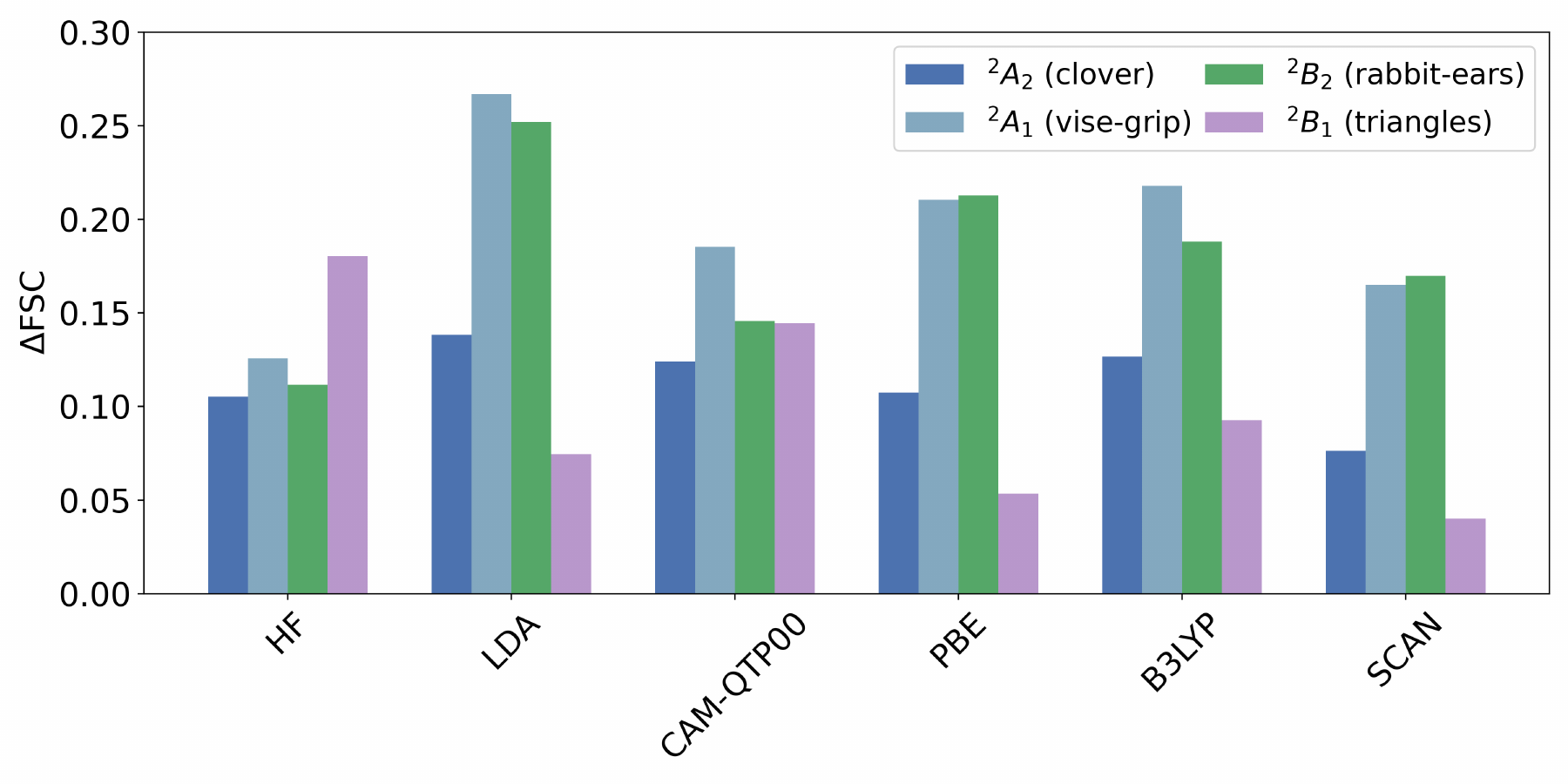}
        \caption{Differences in Fisher-Shannon complexities with respect to AGF2 Dyson orbitals for \ce{O_3} at the NIST experimental geometry. Results are obtained with the cc-pVTZ basis set.}
        \label{fig:o3-dyson-deltaFSC}
    \end{subfigure}
    \caption{Information-theoretic measures for \ce{O_3} Dyson orbitals.}
    \label{fig:o3-dyson}
\end{figure}
% \begin{figure}[H]
%     \centering
%     \includegraphics[width=0.95\linewidth]{figures/o3-dyson-JD-only.pdf}
%     \caption{Jeffreys divergence with respect to AGF2 Dyson orbitals for \ce{O_3} at the NIST experimental geometry. Results are obtained with the cc-pVTZ basis set.}
%     \label{fig:o3-dyson-JD}
% \end{figure}
%orbitals 
\begin{figure}[h]
    \centering
    \begin{subfigure}[b]{0.275\textwidth}
        % \centering
        \includegraphics[trim=100 100 100 100, clip, width=\textwidth]{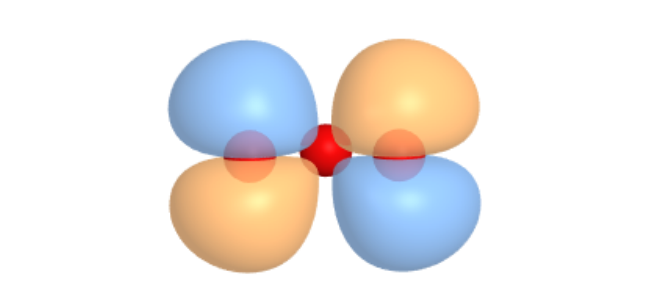}
        \caption{``Clover''}
    \end{subfigure}
    % \hfill
    \begin{subfigure}[b]{0.275\textwidth}
        % \centering
        \includegraphics[trim=100 100 100 100, clip, width=\textwidth]{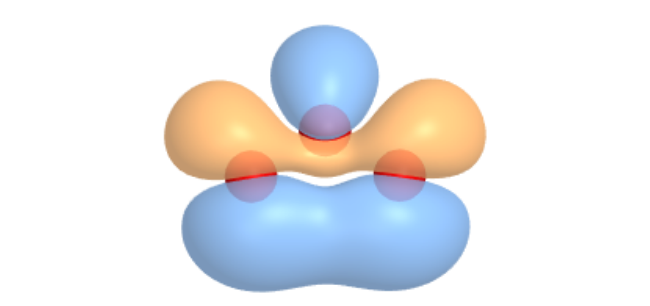}
        \caption{``Vise-grip''}
    \end{subfigure}
 
    \begin{subfigure}[b]{0.275\textwidth}
       % \centering
        \includegraphics[trim=100 100 100 100, clip, width=\textwidth]{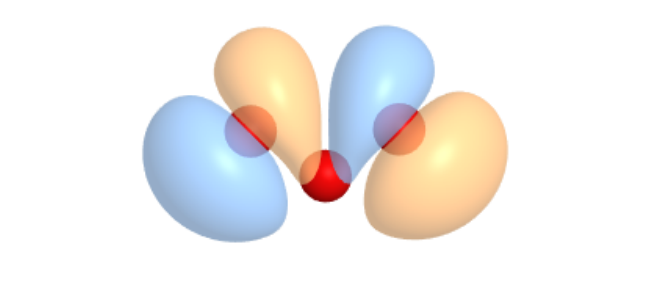}
        \caption{``Rabbit-ears''}
    \end{subfigure}
    % \hfill
    \begin{subfigure}[b]{0.275\textwidth}
        % \centering
        \includegraphics[trim=100 100 100 100, clip, width=\textwidth]{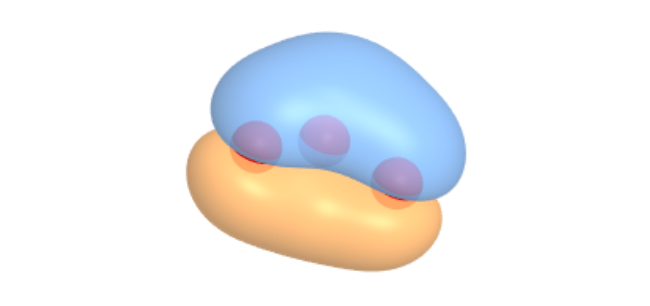}
        \caption{``Triangles''}
    \end{subfigure}

    \caption{Dyson orbitals for \ce{O_3}.}
    \label{fig:o3-orbitals}
\end{figure}
Dyson orbitals (DOs) are single-particle functions defined by the overlap of $N$ and $N\pm1$ wavefunctions and are relevant to photoelectron spectroscopy and the one-electron picture of chemical bonding.\cite{Ortiz1997,Ortiz2012} Their normalizations are determined from the probability factors or pole strengths which may lie between 0 and 1.\cite{Ortiz2020} Since the information-theoretic perspective of DOs is seldom explored, we perform a comparison of densities of MOs and DOs involved in the ionization of an electron. If the pole strength associated with an electron attachment or detachment is close to unity, the corresponding MO from HF\cite{DiazTinoco2019} or even KS-DFT\cite{Duffy1994} can be a good approximation to the DO. 

Final-state cations of ozone have been characterized in prior studies using electron propagator theory\cite{Pawloski2021} and will be further examined here. Figure \ref{fig:o3-dyson-JD} compares the $J$-divergences between different orbitals. Visualizations performed with Vipster v1.17b\cite{vipster} are available in Figure \ref{fig:o3-orbitals}. The canonical HF and KS MOs corresponding to the $^2A_2$ state exhibit the greatest similarity to the DO computed from AGF2 compared to the other orbitals. The Koopmans' theorem result (HF/cc-pVTZ) for this transition is 13.19 eV, which is a close estimate to the experimental value\cite{Katsumata1984} of 13.54 eV, further supporting the viability of approximating the DO with the HF MO. For the remaining transitions, the KS MO densities deviate more substantially from the DOs as  electron correlation effects become more significant. Remarkably, CAM-QTP00 offers the lowest $J$-divergences between its canonical KS MOs and DOs for all the states. Additional data for Shannon and Fisher entropy measures are available in Supporting Figures 6 and 7. 

Changes in the single-particle density shapes or contours are expected when shifting from uncorrelated to correlated orbital theories. Average global and local characteristics of the orbital densities are measured using the Fisher-Shannon complexities (see Supporting Figure 8 and 9). Differences ($\Delta$FSC) in orbital complexity, with respect to AGF2, are depicted in Figure \ref{fig:o3-dyson-deltaFSC}. Nodal and lobe structure contribute to the complexity or electronic organization of the of $\rho(\textbf{r})$ of interest. For example, the FSC measures for orbitals represented by Figures \ref{fig:o3-orbitals}b and \ref{fig:o3-orbitals}c are larger than the other two across all methods employed.

We close this section by noting that the density of the highest occupied molecular orbital is related to the Fukui function.\cite{Melin2005,Melin2007} The Fukui function is central to conceptual DFT\cite{Geerlings2003}, which heralds the density as the fundamental carrier of information for determining chemical reactivity. Thus, ensuring that the $\rho^\mathrm{approx}$ density, both at the orbital and molecular level, is of sufficient quality relative to $\rho^\mathrm{exact}$, should be a primary focus of density functional development. 

% need to plot the bruckner vs canon
% maybe do mesh overlay in iqmol

% \begin{figure}[H]
%     \centering
%     \includegraphics[width=0.95\linewidth]{figures/o3-dyson-deltaFSC.pdf}
%     \caption{Differences in Fisher-Shannon complexities with respect to AGF2 Dyson orbitals for \ce{O_3} at the NIST experimental geometry. Results are obtained with the cc-pVTZ basis set.}
%     \label{fig:o3-dyson-deltaFSC}
% \end{figure}

\subsection{Brueckner Orbitals}

\begin{figure}[H]
    \centering
    \includegraphics[width=0.95\linewidth]{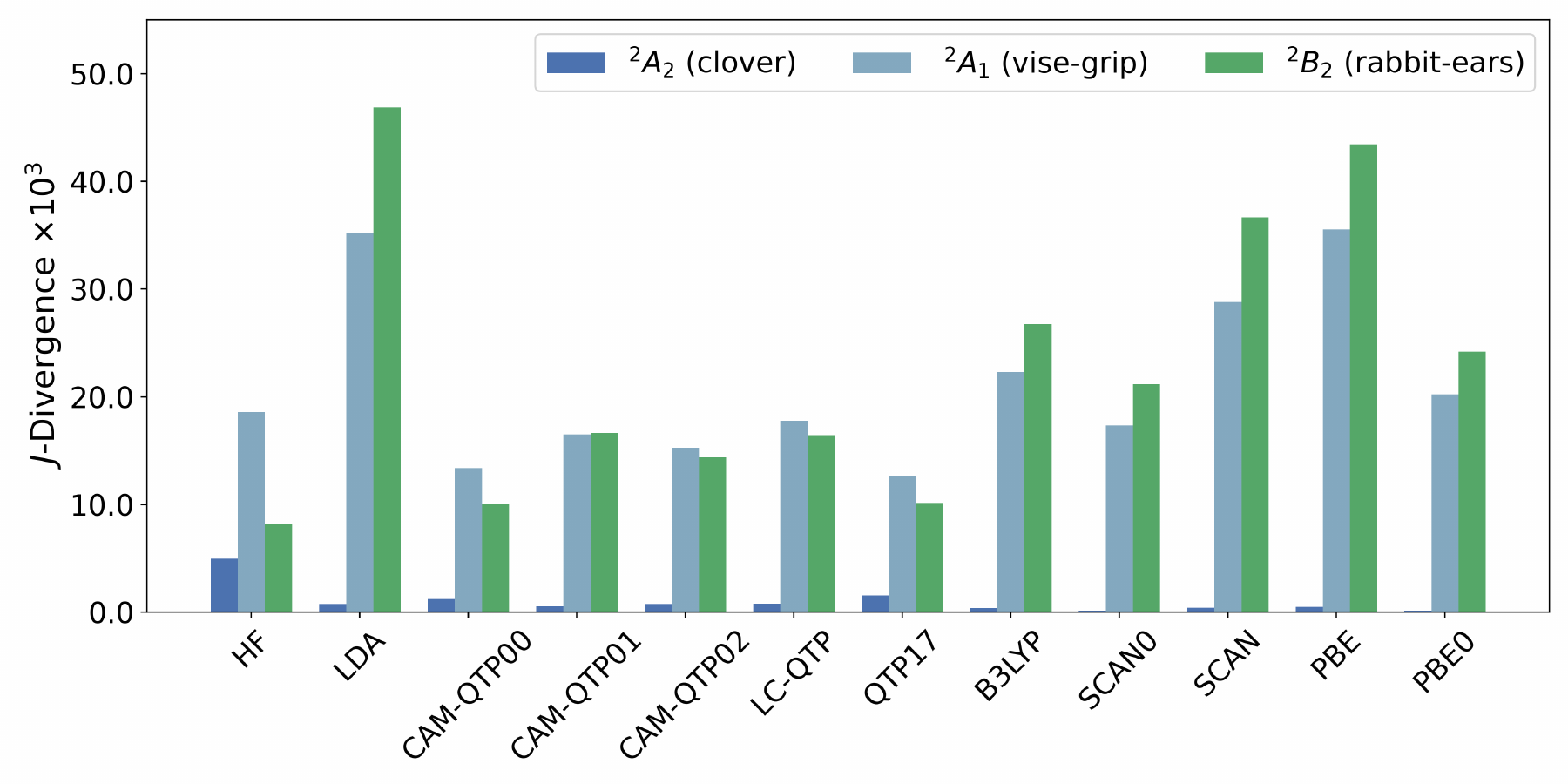}
    \caption{Jeffreys divergence with respect to Brueckner orbitals for \ce{O_3} at the NIST experimental geometry. Results are obtained with the cc-pVTZ basis set.}
    \label{fig:o3-bd-JD}
\end{figure}

Whether or not KS orbitals resemble Brueckner orbitals\cite{Brueckner1954a,Nesbet1958,Stolarczyk1984} (BO) remains an outstanding question.\cite{Jankowski2004,Lindgren2002,Wasilewski2004,Wasilewski2009} We found it worthwhile to take an entropic approach this quandary. We note that BOs and DOs are conceptually similar for systems with a single valence electron above, or a single hole within, a closed-shell configuration.\cite{Lindgren2004} The main difference is that the BOs provide the ingredients to describe a more optimal ground state reference rather than electron binding energies for numerous final-states.\cite{Ortiz2004} Using the notation in the previous section for the highest occupied MOs of \ce{O_3}, calculated $J$-divergences (Figure \ref{fig:o3-bd-JD}) and Shannon entropy differences (Supporting Figure 10) are given. It is interesting to see how the ``clover'' DO, well-described previously by the HF MO, comes into play here. Its BO equivalent becomes nearly indistinguishable from the corresponding KS MO across all DFAs. One might say the KS MO for this particular orbital is superior to the HF MO. All other KS MOs are much less similar to their associated BOs---a trend we observed in comparison to the DOs in \ref{fig:o3-dyson-JD}. Although ground state \ce{O_3} is multi-reference, it is expected that the canonical HF or KS MOs may resemble the BOs only when correlation effects are less important for that orbital. Some amount of non-locality in the XC potential with the inclusion of exact HF exchange appears to lower the $J$-divergences in comparison to the local or semi-local functionals. However, results from hybrid functionals alone was determined to be a weak indicator of similarity to the BOs.\cite{Wasilewski2004} Physical constraints that restore some correlated orbital effects and suppress SIE, such as those that maintain close adherence to Koopmans' theorem, appear  critical for obtaining improved orbital densities. Our results suggest that Kohn-Sham orbitals \textit{can} resemble Brueckner orbitals, but that the similarity varies and is not general. In the case of \ce{O_3} in particular, we find only one orbital of three analyzed to be especially similar in this regard.

%A definitive answer to the query, ``do Kohn-Sham orbitals resemble Brueckner orbitals?'', remains elusive.

\subsection{Dipole Moment of \ce{CO}}

\begin{figure}[H]
    \centering
    \includegraphics[width=0.95\linewidth]{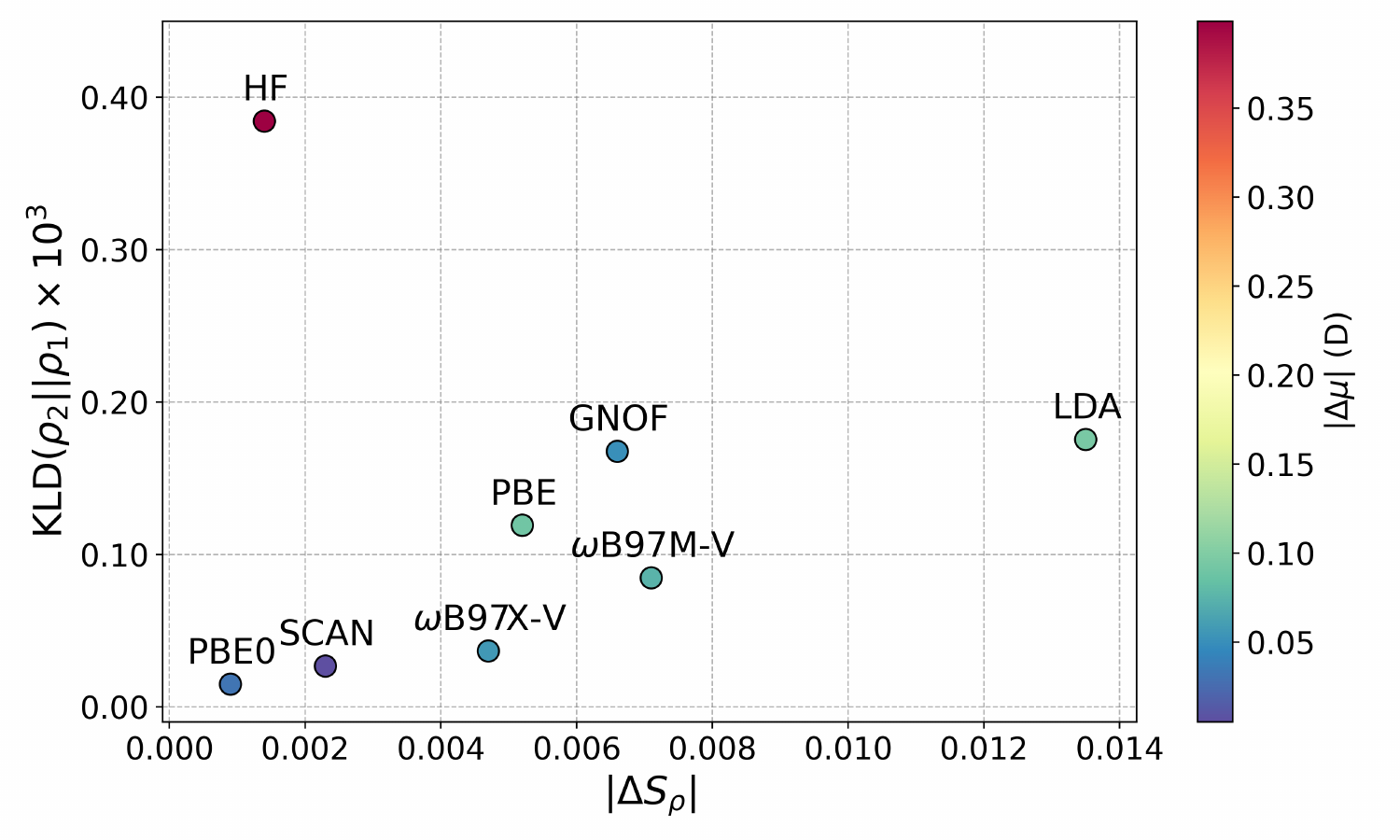}
    \caption{Kullback-Leibler divergences and absolute Shannon entropy differences for \ce{CO} at the NIST experimental geometry. Color-mapped absolute dipole moment differences $|\Delta\mu|$ in Debye (D) are provided. Results are with respect to CCSD and are obtained with the aug-pc-3 basis set\cite{jensen2001a,jensen2002a,jensen2002b} taken from the Basis Set Exchange.\cite{pritchard2019a,feller1996a,schuchardt2007a}}
    \label{fig:co-dipole}
\end{figure}

Dipole moment accuracy is known to vary\cite{Hait2018} across the rungs of Jacob’s ladder\cite{Perdew2001}, so a useful approach in the curation of functionals would be to use entropic measures as key diagnostic tools. To this end, we investigate the relationship between the molecular density information and dipole moment of \ce{CO}. Figure \ref{fig:co-dipole} shows the KLD along with absolute entropy $|\Delta S_\rho|$ and dipole $|\Delta \mu|$ differences, all with respect to CCSD, for HF, GNOF, and an assortment of DFAs. Clearly, HF gives an incorrect sign and magnitude for the dipole vector.\cite{Scuseria1991} SCAN and PBE0 calculations provide the lowest $|\Delta \mu|$ and their densities also confer the lowest KLD and $|\Delta S_\rho|$. The HF $|\Delta S_\rho|$ shows that the accuracy of the dipole moment cannot be consistently predicted using only the position‑space entropy.

Recently, Gubler et al. benchmarked different classes of DFAs to assess the quality of charge densities for molecular properties.\cite{Gubler2025} Interestingly, they find that GGAs, meta-GGAs, and hybrids yield charge densities greater in accuracy than HF and that ``functionals that adhere to theoretical constraints, such as SCAN and r2-SCAN, produced the most consistent and accurate results across all error measures.'' Throughout this work, we have developed similar conclusions, particularly for the likes of SCAN and PBE0, from our information-theoretic perspective.
%yeah uhh the qtp fxnls dont give right dipole sign or too small, i guess qtp01 is ok for dynamic polarizabilities but they dont report raw dipoles https://arxiv.org/pdf/2603.15788  

\subsection{Practical outlook on Information-Constrained DFT}

We have established further the quantitative link between the structure of electron distributions and their inherent uncertainty. This raises the question, how can information theory concepts be used to improve, not just describe, the density? Theoretical developments to include the entropy in the design of kinetic energy functionals\cite{Trickey2011,Gerolin2020} are still nascent. Gadre and co-workers\cite{Gadre1984b,Gadre1988,Gadre1990} have explored a means of refining atomic properties by optimizing basis sets and densities through difference density or relative entropy minimization, with only a small penalty on the total energy. Recapitulating the essence of the entropic approach, we apply KLD minimization over MO coefficients ($\phi$) subject to normalization and second moment constraints $\braket{r^2}$ and $\braket{p^2}$:

\begin{align}
\mathcal{L}(\phi, \lambda_1, \lambda_2, \lambda_3) &= D_{KL}(\phi) 
+ \lambda_1\Big(\langle\phi|\phi\rangle - 1\Big) 
+ \lambda_2\Big(\langle\phi|\hat{r}^2|\phi\rangle - \langle r^2\rangle_\text{ref}\Big) 
+ \lambda_3\Big(\langle\phi|\hat{p}^2|\phi\rangle - \langle p^2\rangle_\text{ref}\Big) . 
\end{align}

This can be solved numerically until the stationary conditions are met. As a simple example, we select the \ce{He} atom and optimize the HF density with a PBE reference (Figure \ref{table:he-refined-to-PBE}). From the number of iterations (see Supporting Figure 11) we see that $\braket{r^2}$, and $\braket{p^2}$ converge much faster than the KLD. In principle, the process can be repeated for multiple orbitals until the $\rho$-derived quantity meets its target value. If a variational approach is desired, the total wavefunction would benefit from position-space KLD minimization over basis sets tuned for density-based properties. Basis sets refined in this manner would have great potential for general-purpose use in SCF calculations. Optimal basis sets have already been constructed for obtaining accurate moments $\braket{p^{-1 \leq k \leq 3}}$ of $\gamma(\textbf{p})$.\cite{Lehtola2012,Lehtola2013} Moreover, optimizations based on mutual information\cite{Sagar2011} may have the prospect of enhancing quantum density methods that involve the on-top pair density.\cite{Burke1998,LiManni2014} We are actively exploring research along these lines.

% dyson is a mix of ov, so harder to minimize kld to zero, lets not include

\begin{table}[H]
\renewcommand{\arraystretch}{1.75}
\setlength{\tabcolsep}{12pt}

\centering
\begin{threeparttable}
\caption{Second moment refinement by relative entropy (KLD) minimization for \ce{He} atom.$^a$}
\label{table:he-refined-to-PBE}
\begin{tabular}{lcccc}
\hline
 & $\braket{r^2}$ & $ \braket{p^2} $ & $\mathrm{KLD}$$^d$  \\
\hline
$\rho_\mathrm{HF}$ & 2.363 & 5.722 & 0.029  \\
$\rho_\mathrm{HF \rightarrow PBE}^\mathrm{opt}$ & 2.369 & 5.777 & 0.000  \\
$\rho_\mathrm{near\text{-}HF}^{\mathrm{ref}}$$^b$ & 2.370 & 5.723 & - \\
$\rho_\mathrm{geminal}^{\mathrm{ref}}$$^c$ & 2.387 & 5.807 & -  \\
\hline
\end{tabular}
\begin{tablenotes}
\footnotesize
\item [$a$] KLD and density refinements are relative to PBE. A core (1s) uncontracted cc-pVTZ basis set is used. Constrained optimizations are performed with SciPy.\cite{scipy,lalee1998,byrd1999} 
\item [$b$] Near-Hartree-Fock\cite{Robertson1986}
\item [$c$] Gaussian geminals\cite{Regier1985}
\item[$d$] $\mathrm{KLD} \times 10^3$
\end{tablenotes}
\end{threeparttable}
\end{table} 

\section{Conclusion} 

The behavior of information entropy measures under various conditions, including bond dissociation, symmetry breaking, ionization, excitation, ensembleization, and confinement was explored with different quantum chemistry methods. The knowledge base we have built in this work certifies that density functionals designed to suppress self-interaction error and to generate good densities relative to those produced by CI or CC methods, are likely to lead to good chemistry. In particular, functionals such as SCAN and PBE, which are both rooted in satisfying exact constraints on the electron density, perform exceptionally well across all properties studied in terms of their capacity to emulate high-level wavefunction-derived densities in an information-theoretic sense. We have characterized the extent that KS MOs coincide with correlated orbitals in order to help address longstanding questions regarding the similarity of KS and Brueckner MOs. Specifically, we found that while KS MOs can sometimes resemble Brueckner ones, the trend is not generalizable, even for different MOs in the same molecule.  Finally, we proposed the idea of systematic property-oriented density refinement with an information-theoretic constrained DFT approach.

Quantities such as the Shannon entropy and Kullback-Leibler divergence can be easily abstracted from virtually any software that performs DFT calculations. These information measures offer a computationally inexpensive route to quantify density differences from approximate mean-field theories against correlated references, providing a practical diagnostic for DFA selection. The addition of these tools in routine computational chemistry analysis will be especially valuable for large systems where KS-DFT remains the only tractable option, and in the curation of novel density-based metrics to aid in selecting density functionals for use with correlated wavefunction theories.\cite{Zamani2026} We have found that the quality of the KS density depends directly on the accuracy of the XC potential, which must be constructed to capture the relevant physics of the chemical problem at hand and/or to satisfy known exact constraints on the density. The performance of SCAN, PBE, and the QTP family of functionals in our analysis of electron densities supports this principle. We hope that our findings point towards new avenues for the design and appraisal of density functionals.

\section*{Acknowledgments}

We thank Md.~Rafi Ul Azam for assistance with validating our fractional occupation SCF code. This research was supported in part by the University of Pittsburgh and the University of Pittsburgh Center for Research Computing and Data through the resources provided. Specifically, this work used the H2P cluster, which is supported by NSF award number OAC-2117681.

\section*{Supporting information}

The Supporting Information is available free of charge at

The following files are available free of charge.
\begin{itemize}
  \item SI.pdf: Supplemental data for the information-theoretic analysis performed in this study.
\end{itemize}

%%%%%%%%%%%%%%%%%%%%%%%%%%%%%%%%%%%%%%%%%%%%%%%%%%%%%%%%%%%%%%%%%%%%%
%% If you are using classical BibTeX rather than biblatex,
%% remove the \printbibliography and uncomment the \bibliograpy one
%%%%%%%%%%%%%%%%%%%%%%%%%%%%%%%%%%%%%%%%%%%%%%%%%%%%%%%%%%%%%%%%%%%%%
 % \printbibliography
 \newpage
 % \subsection{References}
%\bibliography{references.bib}

\begin{mcitethebibliography}{289}
\bibitem[not()]{note1}
Programs that compute the electron momentum density and related properties are
  described in Refs. \citenum{PAREMD} and \citenum{ERKALE}.\relax
\mciteBstWouldAddEndPunctfalse
\mciteSetBstMidEndSepPunct{\mcitedefaultmidpunct}
{}{\mcitedefaultseppunct}\relax
\EndOfBibitem
\end{mcitethebibliography}

\providecommand{\latin}[1]{#1}
\makeatletter
\providecommand{\doi}
  {\begingroup\let\do\@makeother\dospecials
  \catcode`\{=1 \catcode`\}=2 \doi@aux}
\providecommand{\doi@aux}[1]{\endgroup\texttt{#1}}
\makeatother
\providecommand*\mcitethebibliography{\thebibliography}
\csname @ifundefined\endcsname{endmcitethebibliography}
  {\let\endmcitethebibliography\endthebibliography}{}

\newpage

ToC
\rule{0.05in}{1.75in}%
\begin{minipage}[b][1.75in]{3.25in} 
\includegraphics[width=\linewidth,height=1.75in,keepaspectratio]{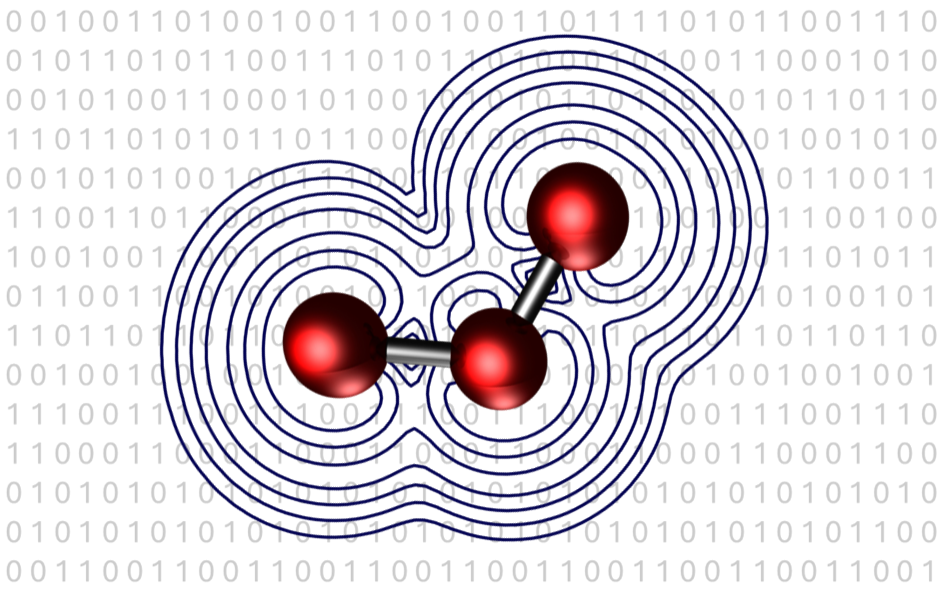}
  \sffamily
  \frenchspacing

%   Some journals require a graphical entry for the Table of Contents. This
%   should be laid out ``print ready'' so that the sizing of the text is correct.

%   The space available depends on the journal: J. Am. Chem. Soc. allows 3.25 in
%   by 1.75 in and requires sanserif text. Some journals want different sizes:
%   you can easily adjust here.
  
%   The two rules either side of the content are there to help judge the height
%   of your material: they may be deleted once not required.
  
\end{minipage}%
\rule{0.05in}{1.75in}

\newpage
\includepdf[pages=-]{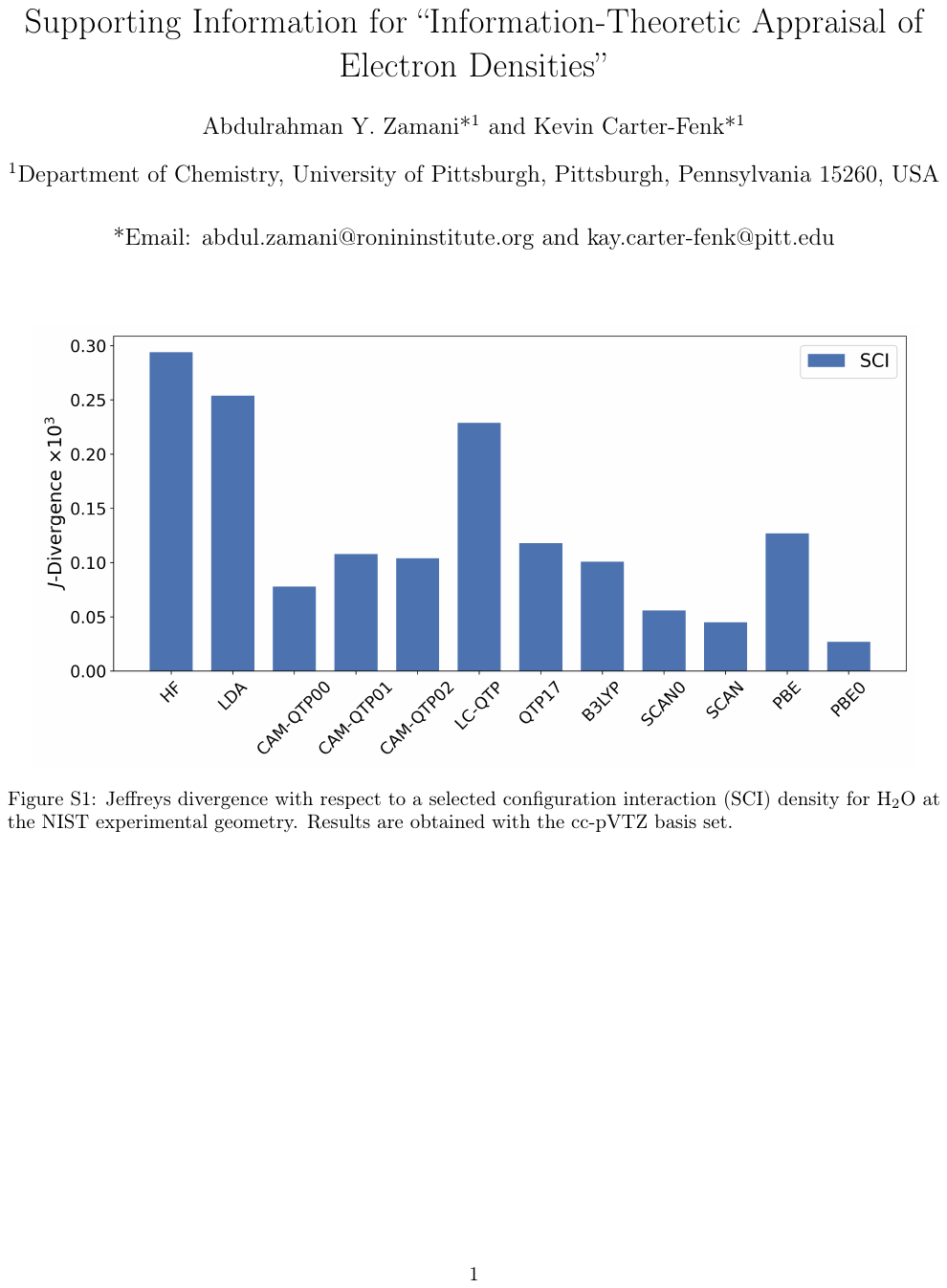}

\end{document}